\documentclass[reqno]{amsart}
\usepackage{amssymb,stmaryrd}
\usepackage{amsfonts}
\usepackage{amstext}
\usepackage{algorithmic}
\usepackage{algorithm}
\usepackage{graphicx}
\usepackage{epstopdf}
\usepackage{color}
\usepackage[all]{xy}
\usepackage{MnSymbol, slashed}

\numberwithin{equation}{section}
\newtheorem{theorem}{Theorem}[section]
\newtheorem{lemma}[theorem]{Lemma}
\newtheorem{proposition}[theorem]{Proposition}

\theoremstyle{definition}

\theoremstyle{remark}

\newcommand{\R}{{\mathbb{R}}}

\newcommand{\C}{{\mathbb{C}}}
\newcommand{\Z}{{\mathbb{Z}}}

\newcommand{\wedgeq}{{\wedge\kern-5pt\cdot}}
\newcommand{\cg}{{\mathfrak g}}

\newcommand{\tens}{\otimes}

\newcommand{\id}{{\rm id}}

\newcommand{\extd}{{\rm d}}
\newcommand{\del}{{\partial}}
\newcommand{\eps}{\epsilon}

  \allowdisplaybreaks

\begin{document}

\title{Quantum Kaluza-Klein theory with $M_2(\C)$}


\author{Chengcheng Liu and Shahn Majid}
\address{Queen Mary, University of London\\
School of Mathematics, Mile End Rd, London E1 4NS, UK}

\email{s.majid@qmul.ac.uk}


\begin{abstract}Following steps analogous to classical Kaluza-Klein theory,  we solve for the quantum Riemannian geometry on $C^\infty(M)\tens M_2(\C)$ in terms of classical Riemannian geometry on a smooth manifold $M$, a finite quantum geometry on the algebra $M_2(\C)$ of $2\times 2$ matrices, and a quantum metric cross term. Fixing a standard form of quantum metric on $M_2(\C)$, we show that this cross term data amounts in the simplest case to a 1-form $A_\mu$ on $M$, which we regard as like a gauge-fixed background field. We show in this case that a real scalar field on the product algebra with its noncommutative Laplacian decomposes on $M$ into two real neutral fields and one complex charged field minimally coupled to $A_\mu$. We show further that the quantum Ricci scalar on the product decomposes into a classical Ricci scalar on $M$, the Ricci scalar on $M_2(\C)$, the Maxwell action $||F||^2$ of $A$ and a higher order $||A.F||^2$ term. Another solution of the QRG on the product has $A=0$ and a dynamical real scalar field $\phi$ on $M$ which imparts mass-splitting to some of the components of a scalar field on the product as in previous work.   \end{abstract}
\maketitle 

\section{Introduction}

The idea that spacetime could be better modelled by noncommutative coordinates due to Planck scale effects has gained currency is recent years, with flat proposals appearing in the 1990s\cite{MaRue,DFR,Hoo}, some of them with quantum Poincar\'e group symmetry\cite{Luk},  and curved models more recently using a new formalism of quantum Riemannian geometry (QRG) as in \cite{BegMa} and references there in. Effectively curved phase space models also came out of quantum groups\cite{Ma:pla}. Moreover, there are several approaches to noncommutative geometry and a mathematically deeper one is that of A. Connes\cite{Con},  particularly via the notion of spectral triple or `abstract Dirac operator'. The key difference in the QRG approach, coming out of experience with quantum groups but not limited to them, is to build up the quantum geometry in layers starting with the `coordinate algebra' $A$, a choice of its extension to an exterior algebra $(\Omega,\extd)$ of `differential forms', a quantum metric $\cg\in \Omega^1\tens_A\Omega^1$ and a compatible torsion free `quantum Levi-Civita connection' (QLC) $\nabla:\Omega^1\to \Omega^1\tens_A\Omega^1$. Some early works in this formalism are \cite{BegMa:gra, Ma:squ,Ma:haw,ArgMa1,ArgMa2} and a recent review for physicists is in \cite{ArgMa4}. One of the key mathematical ingredients is the notion of a bimodule connection \cite{DVM,Mou} which allows for a natural notion of metric-compatibility. With more structure, one can build up all the way to a natural geometric Dirac operator and sometimes arrive at a spectral triple or something close to one\cite{BegMa:spe,BegMa}. 

In this paper, we extend a programme  in \cite{ArgMa4}  to analyse QRGs on tensor product algebras $A=C^\infty(M)\tens A_f$ where $M$ is a classical smooth (pseudo)-Riemannian manifold and $A_f$ is a finite-dimensional possibly noncommutative algebra or `finite quantum geometry'. The idea here is to go back to the standard Kaluza-Klein analysis\cite{KK} for Riemannian geometry on $M\times S^1$, now replacing the coordinate algebra $C^\infty(S^1)$ by some $A_f$. This was full analysed in \cite{ArgMa4} for the case $A_f=\C(\Z_n)$, i.e. replacing $S^1$ by a regular polygon using quantum geometry, but the QRG under reasonable assumptions allowed  no cross-terms in the quantum metric on the product, whereas in the original Kaluza-Klein theory these cross terms encode a $U(1)$ gauge field. Our new result now  for $A_f=M_2(\C)$, the algebra of $2\times 2$ matrices, is to successfully solve for QRGs on the product algebra and to find that the additional freedom in the quantum metric cross term again amounts to a generic background 1-form $A_\mu$ on $M$. These main results, Proposition~\ref{propsol+} and Theorem~\ref{thmmain}, hold for a particular shape of QRG on $M_2(\C)$, which we take to be the standard quantum metric in \cite[Example~8.13]{BegMa} for which a significant 3-dimensional moduli space of QLCs is already known. 

Working in this solution for the background QRG on the product algebra, we next analyse two key questions. Firstly, what does a real scalar field on the product look like if we choose a basis of $M_2(\C)$ and understand such a field as a multiplet of four fields on $M$ with respect to the basis? We find that the canonical QRG Laplacian on the product algebra now appears as the usual Laplacian on $M$ for a certain physical metric $\tilde g$ (differing from the initial classical metric $g$ on $M$ by quadratic $A_\mu A_\nu$ terms).  Two of the four components of a real scalar field do not couple (they are neutral fields) and two of them combine naturally to a single complex field, Proposition~\ref{proplap}.  The second question we ask is how does the QRG Ricci scalar curvature $S$ on the product algebra look in terms of the classical Ricci scalar curvature $\tilde S$ of $\tilde g$, the Ricci curvature of $M_2(\C)$ (which we take to be constant for our chosen QRG on this) and the rest involving $A_\mu$? We find in Theorem~\ref{thmS} that the QRG Ricci scalar on the product is 
\begin{equation*} S_{\rm product}= \tilde S+ {\rm const.} +{1\over 8 h} ||F||^2 + {1\over 8 h^2} ||A.F||^2,\end{equation*}
where the contractions in $||\  ||$ are defined using $\tilde g$,  $F=\extd A$ is the `curvature' of $A$ if this regarded as like a gauge field, and $h$ is a real parameter scale for the quantum metric on $M_2(\C)$.

These main results are in Section~\ref{secKK}, where we solve for the QRG on the product under some algebraic assumptions. {A priori} there are two 1-forms $A_{i\mu}$ entering into the quantum metric cross terms but Proposition~\ref{propsol+} shows that for a QLC to exist they should either both vanish or there is a functional relation between the two so that there is really only one independent 1-form. The proposition then gives the rest of the fields for the QRG in the simplest case of the latter, namely with $A:=A_1$ and $A_2=0$ (there is another similar solution with $A_1=0$ and $A:=A_2$) and the rest of the section studies this solution further. Section~\ref{secA0} provides the analysis for the excluded case $A_1=A_2=0$  where there are more general solutions in which an overall scale factor $h$ in the quantum metric on $M_2(\C)$ need not be constant and instead becomes a dynamical field. Then a scalar field on the product appears as a multiplet on $M$ with some components again having a dynamically generated square-mass, see Proposition~\ref{hscaled}.  Proposition~\ref{propShscaled} moreover shows that the Ricci scalar on the product algebra indeed consists of the original classical scalar curvature of on $M$ and the scalar field Lagrangian for the logarithm $\phi$ of $h$ as the dynamical field. This remarkably follows the same pattern as for $\C(\Z_n)$  in \cite{ArgMa4} even though the quantum geometries $A_f$ are very different. Again, there is no need for a further Higgs field. Section~\ref{secalt} briefly consider how our results change of we take an alternative quantum metric on $M_2(\C)$ for which again the QRG is already known. 

The paper begins with a preliminary Section~\ref{secpre} providing a lightning recap of the formalism of QRG as in \cite{BegMa}. The paper ends with some concluding remarks about directions for further work. In this regard, note that product algebras $C^\infty(M)\tens A_f$ are also studied in Connes approach (under the term `almost commutative' spaces) and one can analyse spectral triples on them, obtaining, for  suitable choices, the standard model of particle physics on $M$ with multiplet structure encoded in $A_f$ and the product spectral triple\cite{Cham}. Gauge fields then appear as allowed `fluctuations' of the spectral triple, and fermionic fields are intrinsic to the analysis since one is working with models of the Dirac operator. By contrast, our work goes onto constructing the QRG with the background field $A_\mu$ as part of the quantum metric, and the only physics we look at is that of a scalar field on the product subject to the noncommutative Klein-Gordon equation there. On the other hand, these two lines are surely not incompatible and spinor fields on the product space in the QRG approach will be considered in a sequel, using Dirac operators on $M_2(\C)$ recently found in \cite{LirMa3,Ma:dir} for the same quantum metric on this as studied here.

\section{Preliminaries}\label{secpre}

We provide a very brief recap of the QRG formalism\cite{BegMa:gra,Ma:squ,Ma:haw,ArgMa1,ArgMa2,BegMa}, which amounts to a coherent framework which includes classical Riemannian geometry when the `coordinate algebra' $A$ is commutative and of the form $C^\infty(M)$, but also the case where this can be noncommutative. We also recap the standard QRG on $M_2(\C)$ from \cite{BegMa}. In the paper we will be solving for QRGs on the tensor product of these two cases. 

Throughout the paper, greek indices $\alpha,\beta,\mu,\nu$ etc. will be used for tensor calculus on the classical manifold $M$ and we may also denote local coordinates $x^\mu$ by $(x,t)$ to remind us that $M$ in our context is intended to be spacetime. We sum over repeated indices unless stated otherwise. 

\subsection{Outline of QRG formalism}\label{secQRG} The formalism works over any field but for our purposes we work over $\C$ and ask that the `coordinate algebra' $A$ is a unital  $*$-algebra. Differentials are formally introduced as a bimodule $\Omega^1$ of 1-forms equipped with a map $\extd:A\to \Omega^1$ obeying the Leibniz rule 
\[ \extd(ab)=(\extd a)b+a\extd b\]
for all $a,b\in A$. This is required to extends to an exterior algebra $(\Omega,\extd)$ generated by $A,\extd A$, with $\extd^2=0$ and $\extd$ obeying the graded-Leibniz rule.  

A quantum metric is $\cg\in \Omega^1\tens_A\Omega^1$ together with a bimodule map inverse $(\ ,\ ):\Omega^1\tens_A\Omega^1\to A$ in the sense 
\[( (\omega,\ )\tens\id)\cg=\omega=\id\tens (\ ,\omega)\]
for all $\omega\in \Omega^1$, and some form of quantum symmetry condition such as $\wedge(\cg)=0$ (and refer to $\cg$ as a generalised quantum metric if no form of symmetry is imposed).  The inversion condition in the classical case just says that the matrices expressing $\cg$ and $(\ ,\ )$ with respect to a basis are mutually inverse. In the quantum case it turns out, however, to force $\cg$ to be central (i.e. to commute with $a\in A$) as shown in \cite{BegMa:gra,BegMa}.  A (left) bimodule connection\cite{DVM,Mou} on $\Omega^1$ is $\nabla:\Omega^1\to \Omega^1\tens_A\Omega^1$ obeying 
\[ \nabla(a.\omega)=a.\nabla\omega+ \extd a\tens\omega,\quad \nabla(\omega.a)=(\nabla\omega).a+\sigma(\omega\tens\extd a)\]
for all $a\in A,\omega\in \Omega^1$, for some `generalised braiding' bimodule map $\sigma:\Omega^1\tens_A\Omega^1\to \Omega^1\tens_A\Omega^1$. The latter, if it exists, is uniquely determined. Classically, we would evaluate the first output of $\nabla$ against a vector field to get the associated covariant derivative, but in QRG we work directly with $\nabla$ itself. A connection is torsion free if the torsion $T_\nabla:=\wedge\nabla-\extd$ vanishes, and metric compatible if 
\[ \nabla \cg=(\nabla\tens \id+ (\sigma\tens\id)(\id\tens\nabla))\cg\]
vanishes. When both vanish, we have a {\em quantum Levi-Civita connection} (QLC). Notice that because $\sigma$ itself depends on $\nabla$,  this is a quadratic condition with the result that a QLC need not be unique or might not exist. The curvature of $\nabla$ is (similarly) defined as 
\[ R_\nabla=(\extd\tens \id- \id\wedge \nabla)\nabla:\Omega^1\to \Omega^2\tens_A\Omega^1.\]
Finally, working over $\C$, we need  $(\Omega,\extd)$ to be a $*$-calculus, $g$ `real' in the sense 
\[ \cg^\dagger=\cg;\quad \dagger:={\rm flip}(*\tens *)\]
and $\nabla$ $*$-preserving in the sense 
\[ \nabla\circ *=\sigma\circ\dagger\circ\nabla,\]
see\cite{BegMa}. These conditions in the classical case and locally with respect to real coordinates $x^\mu$ say that the tensor for $\cg$ is real (when combined with quantum symmetry) and that the Christoffel symbols are real. 

Also of interest will be the QRG Laplacian
\[ \Delta:=(\ ,\ )\nabla\extd: A\to A\]
which in the classical case with $\nabla$ the Levi-Civita connection recovers the Laplace-Beltrami operator, and the Ricci tensor. A `working definition' of the latter  (but just based on copying classical formulae)  is to assume a bimodule lifting map $i:\Omega^2\to \Omega^1\tens\Omega^1$ and then take a trace,
\[ {\rm Ricci}=((\ ,\ )\tens \id)(\id\tens i\tens\id)(\id\tens R_\nabla)\cg\in \Omega^1\tens_A\Omega^1,\quad S=(\ , ){\rm Ricci}\]
where $S$ is the Ricci scalar curvature. Note that the classical cases of ${\rm Ricci}$ and $S$ in the natural conventions here as  $-{1/2}$ of their usual values\cite{BegMa}. If $i\circ *=-\dagger\circ i$ along with our other `reality' assumptions then $S$ will be self-adjoint, but we do not necessarily impose this. Likewise $\Delta$ will commute with $*$ if in addition $(\ ,\ )\sigma=(\ , \ )$. These are both strong conditions and we do not impose them directly.

\subsection{Reference QRG on $M_2(\C)$}\label{QRGM2}

We let $\Omega^1(M_2(\C))$ be the standard 2D in \cite{BegMa}, which is given there terms of basis of central 1-forms $s,t$ with $s^*=-t$ and $\Omega(M_2(\C))$ with $s^2=t^2$, $st=ts$, see \cite[Chap.~1]{BegMa}.  We prefer here to work the real self-adjoint basis
\[ s^1={(s+t)\over\imath},\quad  s^2=s-t\]
where now\cite{LirMa3} 
\[ s^1\wedge s^1=s^2\wedge s^2=\imath{\rm Vol},\quad s^1\wedge s^2=s^2\wedge s^1=0\]
with differentials 
\[ \extd \sigma^1=\sigma^3s^2,\quad \extd\sigma^2=-\sigma^3 s^1,\quad \quad \extd \sigma^3=\sigma^2 s^1-\sigma^1 s^2,\quad \extd s^i=-\sigma^i{\rm Vol},\quad {\rm Vol}=-\imath s^1\wedge s^1 \]
The partial derivatives with respect to this basis are defined by $\extd f=(\del_i f)s^i$ (sum over $i=1,2$) and are derivations on $M_2(\C)$ given by
\[ \del_i\sigma^j=-\eps_{ijk}\sigma^k\]
for $i=1,2$ and $j,k\in\{1,2,3\}$. We will mostly be interested in the standard QRG with quantum metric (-2 times)  $s \tens s+t\tens t$ which  now appears as 
\[\cg_{M_2}=s^1\tens s^1 -  s^2\tens s^2\]
with a `Minkowski signature' metric.  The QLC is far from unique with the most well-studied but the simplest case \cite{BegMa:cur,LirMa3} being the 1-parameter family with
\begin{align*}\sigma(s^i\tens s^j)&=\begin{cases}s^i\tens s^j & {\rm if\ }i\ne j\\ -s^{\bar i}\tens s^{\bar i}-2\imath\rho s^{\bar i}\tens s^i & {\rm if\ }i=j\end{cases}\\
\nabla s^i&= {\imath\over 2}\sigma^i(s^1\tens s^1+s^2\tens s^2)+ {\imath\over 2}\eps_{ij}\sigma^j(s^2\tens s^1-s^1\tens s^2)-\rho\sigma^i s^{\bar i}\tens s^i\\
R_{\nabla}s^i&=   -\imath(1+\rho^2 ){\rm Vol}\tens s^i,\quad {\rm Ricci}=-{1\over 2}(1+\rho^2)(s^1\tens s^1+s^2\tens s^2),\quad S=0,
\end{align*}
where $\bar 1=2$ and $\bar 2=1$ and the  Ricci tensor is for the  canonical symmetric lift
\[ i({\rm Vol})=\frac{1}{2\imath}(s^1\tens s^1+ s^2\tens s^2).\]
Here the parameter $\rho$ is required to be imaginary for the $*$-preserving property. The 
Laplacian in this 1-parameter case is 
\begin{align*}
\Delta_{M_2}f&= f_1\sigma^1- f_2\sigma^2
\end{align*}
if we write $f=f_01+ f_i\sigma_i$ with a sum over $i=1,2,3$. We see that $1,\sigma_3$ are zero modes and $\Delta$ is diagonal on $\sigma^1,\sigma^2$. We will use this QRG in Section~\ref{secA0}. 

In the main part of the paper, however, we will need to start off with the full 3-parameter QLCs on $M_2(\C)$ in \cite[Exercise~8.3]{BegMa} with parameters $\rho,u,v$, of which the above is the special case $u=v=0$. We convert this our self-adjoint basis and introduce Christoffel symbols defined by 
\[ \nabla s^k=\gamma^k_{ijp}\sigma^p s^i\tens s^j;
\quad \gamma^k_{ijp}\in \C.\]
Then the 3-parameter QLC appears after some calculation as 
\begin{align*}
\gamma^1_{221}&=\gamma^1_{212}=-\gamma^2_{211}=\imath-\gamma^1_{111},\quad\gamma^1_{122}=-\gamma^2_{121}=-\gamma^2_{112}=\frac{\imath}{2}(u+v)-\gamma^1_{111},\\
\gamma^2_{222}&=\frac{\imath}{2}(u+v+2)-\gamma^1_{111},\quad\gamma^1_{112}=\gamma^1_{222}=\gamma^2_{111}=\gamma^2_{221}=\gamma^2_{212}=\gamma^1_{121},\\
\gamma^1_{211}&=\gamma^1_{121}+\frac{u-v}{2}-\rho,\quad \gamma^2_{122}=\gamma^1_{121}-\frac{u-v}{2}-\rho,\\
\gamma^1_{111}&=\frac{\imath}{8\rho}((v+u-2)(v-u+2\rho)+8\rho),\quad \gamma^1_{121}=\frac{u+v}{8\rho}(2-u-v)
\end{align*}
and the reality condition for the QLC to be $*$-preserving is 
\[ \bar u=v,\quad \bar v=u,\quad \bar \rho=-\rho.\]

\begin{lemma} For the 3-parameter QLCs on $M_2\(C)$, the Riemann curvature, Ricci tensor and Ricci scalar are
\begin{align*}
R_{\nabla}s^1&=-\frac{\imath}{2}((\rho-u)^2+(\rho+v)^2-2(u+v-1)){\rm Vol}\tens s^1-\frac{1}{2}(u-v)(u+v-2){\rm Vol}\tens s^2,\\
R_{\nabla}s^2&=\frac{1}{2}(u-v)(u+v-2){\rm Vol}\tens s^1-\frac{\imath}{2}((\rho+u)^2+(\rho-v)^2-2(u+v-1)){\rm Vol}\tens s^2,\\
{\rm Ricci}&=-\frac{1}{4}((\rho-u)^2+(\rho+v)^2-2(u+v-1))s^1\tens s^1 +\frac{\imath}{4}(u-v)(u+v-2)(s^1\tens s^2-s^2\tens s^1)\\
&\quad -\frac{1}{4}((\rho+u)^2+(\rho-v)^2-2(u+v-1))s^2\tens s^2,\\
S&=\rho (u-v),
\end{align*}
and the Laplacian is 
\begin{align}\label{delta2m}
\Delta_{M_2}f&=-\frac{(u+v-2)}{4\rho}((u-v+2\rho)f_1\sigma^1+ (u-v-2\rho)f_2\sigma^2+2(u-v)f_3\sigma^3)
\end{align}
if we write $f=f_01+ f_i\sigma_i$ with a sum over $i=1,2,3$. 
\end{lemma}
\proof We used the definitions in Section~\ref{secQRG}, with calculus and lifting map as stated above. Since the basis $s^i$ is central, all calculations can be reduced to working with tensors of coefficients (which were then computed with Mathematica.)\endproof

We see that the Ricci scalar curvature $S$ is a constant (a multiple of $1\in M_2(\C)$) and real in the $*$-preserving case. Both the QRGs  on $M_2(\C)$ discussed are dimensionless, but we are free to scale the quantum metric by a dimensionful constant $h$, which does not, however, change $\nabla$. 

\section{Kaluza-Klein model and scalar fields on $C^\infty(M)\tens M_2(\C)$}\label{secKK}

We follow the line recently introduced in \cite{ArgMa4} but with $\C(\Z_n)$ replaced by $M_2(\C)$. Thus, we analyse the QRG on the tensor product algebra  $A=C^\infty(M)\tens M_2(\C)$ where $M$ is a classical manifold. We let $\Omega(M)$ be the classical exterior algebra on $M$ (we will be mainly concerned with 1-forms) and let central $s^i$ be the basis of $\Omega^1(M_2(\C))$ in Section~\ref{QRGM2}.

For the differential calculus on the product, it is natural to take the graded tensor product exterior algebra $\Omega(A)=\Omega(M)\underline\tens \Omega(M_2(\C))$, or explicitly
\begin{align*}
(\omega\tens \eta)(\omega'\tens \eta')=(-1)^{\deg(\eta)\deg(\omega')}\omega \omega'\tens \eta \eta'
\end{align*}
for all $\omega,\omega'\in \Omega(M)$ and $\eta,\eta'\in \Omega(M_2(\C))$. For the sake of calculations, we will work with local coordinates $x^\mu$ on $M$ and use our basis $s^i$ on $M_2(\C)$. It is easy to see that (locally)  when viewed in $\Omega^1(A)$, the $\extd x^\mu,s^i$ together are a central basis over $A$ and mutually anticommute, the  $\extd x^\mu$ remain Grassmann as  usual and the $s^i$ retain their wedge product from Section~\ref{QRGM2}. For example, 
\begin{align*}(\extd x^\mu\tens 1)(f\tens m)&= \extd x^\mu f\tens m=f\extd x^\mu \tens m= (f\tens m)(\extd x^\mu\tens 1),\\
(1\tens s^i)(f\tens m)&= f\tens s^i m=f\tens m s^i=(f\tens m)(1\tens s^i),\\
 (\extd x^\mu\tens 1)(1\tens s^i)&=\extd x^\mu\tens s^i=-(1\tens s^i)(\extd x^\mu\tens 1),\end{align*}
etc. Henceforth, we identify $\extd x^\mu\equiv\extd x^\mu\tens 1,s^i\equiv 1\tens s^i,f\equiv f \tens 1, m\equiv 1 \tens m$ for $f\in C^\infty(M)$ and $m\in M_2(\C)$ when viewed in the product exterior algebra and freely use these properties.  Similarly, one can check that the classical antisymmetric lift
\[  i(\extd x^\mu\wedge\extd x^\nu)={1\over 2} (\extd x^\mu\tens \extd x^\nu- \extd x^\nu\tens\extd x^\mu)\]
extends by the same formula when we work in the product algebra and also when exactly one of the $\extd x^\mu,\extd x^\nu$ are replaced by $s^i$. Also recall that we sum over repeated indices unless stated otherwise.

\begin{lemma}\label{lemg} The most general quantum metric on the tensor product algebra $C^\infty(M)\tens M_2(\C)$ has the form
\[\cg=g_{\mu\nu}(x,t)\extd x^\mu\tens \extd x^\nu+A_{i\mu}(x,t) (s^i\tens \extd x^\mu+\extd x^\mu\tens s^i)+h_{ij}(x,t) s^i\tens s^j,\]
where $g_{\mu\nu}$ are symmetric, $h_{11}+h_{22}=0$ and all coefficients are functions on spacetime $M$. Here $i,j\in\{1,2\}$. 
\end{lemma}
\proof For $M_2(\C)$, we use the Pauli basis $\sigma^a$  for $a=0,1,2,3$ with $\sigma^0:=\id$. Then {\em a priori}, the most possible general quantum metric is 
\begin{align*}
\cg=g_{a\mu\nu}(x,t)\sigma^a \extd x^\mu\tens \extd x^\nu+A_{a\mu i}(x,t)\sigma^a \extd x^\mu\tens s^i+A_{ai\mu}(x,t)\sigma^a s^i\tens \extd x^\mu+h_{aij}(x,t)\sigma^a s^i\tens s^j
\end{align*}
where $g_{a\mu\nu},A_{a\mu i}(x,t),A_{ai\mu}(x,t),h_{aij}\in C^\infty(M)$. 
Because the quantum metric has to be central, one needs for all $a\in A$,  
\begin{align*}
[\cg,a]&=[g_{a\mu\nu}(x,t)\sigma^a, a]\extd x^\mu\tens \extd x^\nu+[A_{a\mu i}(x,t)\sigma^a, a]\extd x^\mu\tens s^i\\
&\quad +[A_{ai\mu}(x,t)\sigma^a,a] s^i\tens \extd x^\mu +[h_{aij}(x,t)\sigma^a,a] s^i\tens s^j=0
\end{align*}
which means 
 \[g_{a\mu\nu}(x,t)\sigma^a=g_{\mu\nu}(x,t),
\quad A_{a\mu i}(x,t)\sigma^a=A_{\mu i}(x,t),\]\[ A_{ai\mu}(x,t)\sigma^a=A_{i\mu}(x,t),\quad h_{aij}(x,t)\sigma^a=h_{ij}(x,t),\]
for $g_{\mu\nu},A_{\mu i}(x,t),A_{i\mu}(x,t),h_{ij}\in C^\infty(M)$, i.e. only multiples of $1\in M_2(\C)$ as the centre of this algebra. 

Now we impose the quantum symmetry condition for the quantum metric in the form $\wedge(\cg)=0$,
\begin{align*}
&\wedge(\cg)=g_{\mu\nu}(x,t)\extd x^\mu\wedge \extd x^\nu+A_{\mu i}(x,t) \extd x^\mu\wedge s^i+A_{i\mu}(x,t) s^i\wedge \extd x^\mu+h_{ij}(x,t) s^i\wedge s^j\\
&=g_{\mu\nu}(x,t)\extd x^\mu\wedge \extd x^\nu+A_{\mu i}(x,t) \extd x^\mu\wedge s^i+A_{i\mu}(x,t) s^i\wedge \extd x^\mu+[h_{11}(x,t)+h_{22}(x,t)]\imath {\rm Vol}\\
&=g_{\mu\nu}(x,t)\extd x^\mu\wedge \extd x^\nu+(A_{i\mu}(x,t)-A_{\mu i}(x,t)) s^i\wedge \extd x^\mu+[h_{11}(x,t)+h_{22}(x,t)]\imath {\rm Vol}\\
&=0
\end{align*} 
which implies the stated conditions on $A_{\mu i}$ and $h_{ij}$.  \endproof

In particular, we can consider $g_{\mu\nu}$ as a classical (pseudo) Riemannian metric on $M$. Strictly speaking it does not need to be invertible but only the larger matrix corresponding to $\cg$, but it is convenient to assume that $g$ is separately invertible as part of the generic solution.  From now on, we assume for simplicity that if $\sigma$ exists, it obeys the following `flip assumption':
\begin{equation}\label{condi}\sigma(\extd x^\mu\tens s^i)=s^i\tens\extd x^\mu.\end{equation}
This says that we are not too far from the classical case where $\sigma$ is always just a flip map, and was similarly assumed in order to simplify the analysis in \cite{ArgMa4}.

\begin{proposition}\label{tornabla}
Under the assumption (\ref{condi}), if $\nabla$ is a torsion free bimodule connection then it has the form
\begin{align*}
\nabla \extd x^\mu=-\Gamma^\mu_{\alpha\beta}\extd x^\alpha\tens \extd x^\beta+B^\mu_{i \alpha}(\extd x^\alpha\tens s^i+ s^i\tens \extd x^\alpha)+C^\mu_{ij} s^i\tens s^j,\\
\nabla s^k=D^k_{\alpha\beta}\extd x^\alpha\tens \extd x^\beta+E^k_{i\alpha }(\extd x^\alpha\tens s^i+ s^i\tens \extd x^\alpha)+\gamma^k_{ij} s^i\tens s^j
\end{align*}
where $\Gamma^\mu_{\alpha\beta}, D^k_{\alpha\beta}$ are symmetric for the subscripts, $\Gamma^\mu_{\alpha\beta},B^\mu_{i\alpha},C^\mu_{ij}\in C^\infty(M)$,  $D^k_{\alpha\beta},E^k_{i\alpha},\gamma^k_{ij}\in C^\infty(M)\tens M_2(\C)$ but do not have the $\sigma^3$ component, and 
\[ C^\mu_{11}+C^\mu_{22}=0,\quad \gamma^k_{11}+\gamma^k_{22}=\imath \sigma^k.\]Here $i,j,k\in\{1,2\}$.
\end{proposition}
\proof
{\em A priori}, the most general form of connection is
\begin{align*}
\nabla \extd x^\mu=-\Gamma^\mu_{\alpha\beta}\extd x^\alpha\tens \extd x^\beta+A^\mu_{\alpha i}\extd x^\alpha\tens s^i+B^\mu_{i \alpha} s^i\tens \extd x^\alpha+C^\mu_{ij} s^i\tens s^j,\\
\nabla s^k=D^k_{\alpha\beta}\extd x^\alpha\tens \extd x^\beta+E^k_{i\alpha}\extd x^\alpha\tens s^i+F^k_{\alpha i} s^i\tens \extd x^\alpha+\gamma^k_{ij} s^i\tens s^j.
\end{align*}
Requiring the torsion free condition  $\wedge\nabla-\extd=0$, we get
\[\Gamma^\mu_{\alpha\beta}=\Gamma^\mu_{\beta\alpha},\quad A^\mu_{\alpha i}=B^\mu_{i \alpha},\quad C^\mu_{11}+C^\mu_{22}=0,\]
\[D^k_{\alpha\beta}=D^k_{\beta\alpha},\quad E^k_{i\alpha}=F^k_{ \alpha i},\quad \gamma^k_{11}+\gamma^k_{22}=\imath \sigma^k,\]
so a torsion free connection has the form stated.

From $\nabla (e^\alpha a)=\nabla (a e^\alpha)$ for $e^i$ denoting either $\extd x^\mu$ or $s^i$, we also deduce $\sigma(e^\alpha\tens\extd a)=[a,\nabla e^\alpha]+\extd a\tens e^\alpha$ which in particular means
\begin{align*}
\sigma(e^\alpha\tens \extd x^\mu)=[x^\mu,\nabla e^\alpha]+\extd x^\mu\tens e^\alpha=\extd x^\mu\tens e^\alpha
\end{align*}
so that $\sigma(e^\alpha\tens \extd x^\mu)$ is flip, and
\begin{align}\label{sigmaes}
\begin{split}
\sigma(e^\alpha\tens s^j)&=\sigma(e^\alpha\tens (-1)^j\sigma^3\extd \sigma^{\bar j})=(-1)^j\sigma^3\sigma(e^\alpha\tens \extd \sigma^{\bar j})\\
&=(-1)^j\sigma^3([\sigma^{\bar j},\nabla e^\alpha]+\extd \sigma^{\bar j}\tens e^\alpha)\\
&=(-1)^j\sigma^3[\sigma^{\bar j},\nabla e^\alpha]+s^j\tens e^\alpha.
\end{split}
\end{align}
Here, we use $s^j=(-1)^j\sigma^3\extd \sigma^{\bar j}$ in the first step. Therefore according to (\ref{sigmaes}), assumption (\ref{condi}) gives
\[[\sigma^j,\nabla \extd x^\mu]=0\]
for $j=1,2$.  It then follows that
\[[\sigma^3,\nabla\extd x^\mu]=-\imath[\sigma^1\sigma^2,\nabla \extd x^\mu]=0\]
and hence $\Gamma^\mu_{\alpha\beta},B^\mu_{i\alpha},C^\mu_{ij}$ are in the centre of $M_2(\C)$. One can also find that
\begin{align*}
[\sigma^3,\nabla e^\alpha]&=[-\imath\sigma^1\sigma^2,\nabla e^\alpha]=-\imath\sigma^1[\sigma^2,\nabla e^\alpha]-\imath [\sigma^1,\nabla e^\alpha]\sigma^2\\
&=\imath\sigma^1\sigma^3(\sigma(e^\alpha\tens s^1)-s^1\tens e^\alpha)-\imath\sigma^3(\sigma(e^\alpha\tens s^2)-s^2\tens e^\alpha)\sigma^2\\
&=\sigma^2(\sigma(e^\alpha\tens s^1)-s^1\tens e^\alpha)+\sigma^1 (\sigma(e^\alpha\tens s^2)-s^2\tens e^\alpha)
\end{align*}
 under assumption (\ref{condi}).

Next, because our connection is assumed to be a bimodule connection, $\sigma$ is required to be a bimodule map so that $a\sigma(e^\alpha\tens e^\beta)=\sigma(ae^\alpha\tens e^\beta)=\sigma(e^\alpha\tens e^\beta a)=\sigma(e^\alpha\tens e^\beta)a$ for any $a\in A$. In particular, for $\sigma(s^i\tens s^j)$ from (\ref{sigmaes}), this gives
\[[a,\sigma^3[\sigma^j,\nabla s^k]]=0\]
or
\begin{align}\label{DEg}
[a,\sigma^3[\sigma^m,D^k_{\alpha\beta}]]=0,\quad [a,\sigma^3[\sigma^m,E^k_{i\alpha}]]=0,\quad [a,\sigma^3[\sigma^m,\gamma^k_{ij}]]=0
\end{align}

where $m=1,2$. We can rewrite $D^k_{\alpha\beta}=D^k_{\alpha\beta n}\sigma^n, E^k_{i\alpha}=E^k_{i\alpha n}\sigma^n, \gamma^k_{ij}=\gamma^k_{ijn}\sigma^n$ for $D^k_{\alpha\beta n},E^k_{i\alpha n},\gamma^k_{ijn}\in C^\infty(M),\sigma^n\in M_2(\C),n=0,1,2,3$. Then the last equation in (\ref{DEg}) gives
\[\gamma^k_{ijn}[a,\sigma^3[\sigma^m,\sigma^n]]=0\]
When $m=1$, we deduce
\[\gamma^k_{ij1}[a,\sigma^3[\sigma^1,\sigma^1]]+\gamma^k_{ij2}[a,\sigma^3[\sigma^1,\sigma^2]]+\gamma^k_{ij3}[a,\sigma^3[\sigma^1,\sigma^3]]=0\]
or
\[\gamma^k_{ij3}[a,\sigma^1]=0.\]
For this to hold for all $a\in A$, we need $\gamma^k_{ij3}=0$. When $m=2$, it also gives that $\gamma^k_{ij3}=0$. Similarly, the first and second equations in (\ref{DEg}) give $D^k_{\alpha\beta 3}=0$ and $E^k_{i\alpha 3}=0$.
\endproof

We still need to show that we can actually obtain a bimodule connection in this way:

\begin{lemma}\label{lemnabla} Every $\nabla$ of the form in Proposition~\ref{tornabla} indeed is a bimodule connection with $\sigma$ the flip on the basis involving $\extd x^\mu$ and
\begin{align}\label{ssij}
\sigma(s^i\tens s^j)=s^j\tens s^i+ 2\imath (D^i_{\alpha\beta j}\extd x^\alpha\tens \extd x^\beta+E^i_{k\alpha j}(\extd x^\alpha\tens s^k+ s^k\tens \extd x^\alpha)+\gamma^i_{klj}s^k\tens s^l]),
\end{align} 
where $i,j,k,l\in\{1,2\}$ and $D^i_{\alpha\beta j},E^i_{k\alpha j},\gamma^i_{klj}\in C^\infty(M)$. \end{lemma}
\proof 
Following our assumptions, let $\sigma$ be the flip on the basis if one of elements is $\extd x^\mu$ and let
\begin{align*}
&\sigma(s^i\tens s^j)\\
=& s^j\tens s^i + (-1)^j\sigma^3[\sigma^{\bar j},D^i_{\alpha\beta f}\sigma^f\extd x^\alpha\tens \extd x^\beta+E^i_{k\alpha f}\sigma^f(\extd x^\alpha\tens s^k+ s^k\tens \extd x^\alpha)+\gamma^i{}_{klf}\sigma^f s^k\tens s^l]\\
=&s^j\tens s^i + 2\imath [D^i_{\alpha\beta j}\extd x^\alpha\tens \extd x^\beta+E^i_{k\alpha j}(\extd x^\alpha\tens s^k+ s^k\tens \extd x^\alpha)+\gamma^i_{klj}s^k\tens s^l]\\
=&\sigma^{ij}{}_{\alpha\beta}e^\alpha\tens e^\beta
\end{align*}
for some $D,E,\gamma$ as above and here $f=0,1,2$. And one should also notice that in the third line $j\in\{1,2\}$. This shows that $\sigma^{ij}{}_{\alpha\beta}$ is in the centre of $M_2(\C)$. Now, if $\sigma^{ij}{}_{\alpha\beta}$ is in the centre
of $M_2(\C)$ and the rest of $\sigma$ is the flip map under our assumptions then one can see that $\sigma$ is a bimodule map when extended to $\sigma(a e^\alpha \tens e^\beta b)=a\sigma(e^\alpha\tens e^\beta)b$. Here
\[\sigma(e^\alpha a\tens e^\beta)=\sigma(ae^\alpha\tens e^\beta)=a\sigma(e^\alpha\tens e^\beta)=\sigma(e^\alpha\tens e^\beta)a=\sigma(e^\alpha\tens e^\beta a)=\sigma(e^\alpha\tens ae^\beta)\]
so this gives a well-defined map on $\Omega^1\tens_A\Omega^1$. Hence
$\nabla$ in the proposition is indeed a bimodule connection.  \endproof

Proposition~\ref{tornabla}  is therefore the most general form of torsion free bimodule connection on the product under our assumption (\ref{condi}). We now analyse which of these are QLC's for the quantum metric in Lemma~\ref{lemg}. We already saw that $g_{\alpha\beta}$ and $\Gamma^\rho_{\alpha\beta}$ are symmetric in their lower indices and consider this part of the data as respectively as a classical metric  and a  torsion free connection, which we denote $\nabla_\mu$, on $M$. The following lemma then reduces everything to tensor calculus on $M$  and internal latin indices $i,j,k,l,m,p,q$ etc., relating to $M_2(\C)$.

\begin{lemma}\label{lem_metric} We assume  (\ref{condi}), a quantum metric as in Lemma~\ref{lemg} with $h_{11}+h_{22}=0$, and a torsion free bimodule connection $\nabla$ as in Proposition~\ref{tornabla} with and 
\[\gamma^k_{11}+\gamma^k_{22}=\imath \sigma^k,\quad C^\mu_{11}+C^\mu_{22}=0\]
as before. Then $\nabla$ is a QLC if and only if 
\begin{align*}
&A_{m\alpha} A_{n\beta} A_{i\nu}h^{mj}(h^{il}\gamma^n_{jlk}+h^{nl}\gamma^i_{jlk}+2\imath h^{pl}\gamma^n_{jlq}\gamma^i_{pqk})=0,\\
&A_{m\alpha}A_{n\beta}h^{ml}(\gamma^n_{lik}+h_{ji}h^{jp}\gamma^n_{lpk}+2\imath h^{sr}h_{sj}\gamma^n_{lrp}\gamma^j_{pik})=0,\\
&A_{m\alpha}A_{n\beta}(h^{mi}\gamma^n_{ipk}+\imath h^{iq}\gamma^n_{qpj}\gamma^m_{ijk})=0,\\
&A_{i\alpha}A_{j\beta}(h^{in}\gamma^j_{nl k}-h^{jn}\gamma^i_{nl k})=0,\\
&A_{m\alpha}(\gamma^m_{ijk}+h_{nj}h^{nl}\gamma^m_{lik}+2\imath h^{lp}h_{ln}\gamma^m_{piq}\gamma^n_{qjk})=0,\\
&A_{m\alpha}(\gamma^m_{ijk}+\gamma^m_{ji k}+2\imath \gamma^p_{ijq}\gamma^m_{pq k})=0,\\
&h_{ji}\gamma^j_{mnk}+h_{nj}\gamma^j_{mik}+2\imath h_{pj}\gamma^p_{mnq}\gamma^j_{qik}=0
\end{align*}
and the differential conditions
\begin{align}\label{BC2}
&\nabla_\alpha g_{\beta\nu}+2\imath A_{p\alpha}A_{q\beta}h^{ik}h^{pl}\gamma^q_{lkj}(g_{\nu\mu}C^\mu_{ij}+A_{m\nu}\gamma^m_{ij0}-\partial_\nu h_{ij})\nonumber\\
&+A_{p\nu}h^{pi}(g_{\alpha\mu}B^\mu_{i\beta}+A_{q\alpha}h^{qn}(g_{\beta\mu}C^\mu_{ni}+A_{m\beta}\gamma^m_{ni0}-\partial_\beta h_{ni}-A_{n\mu}B^\mu_{i\beta})-\nabla_\alpha A_{i\beta})\nonumber\\
&+A_{p\beta}h^{pi}(g_{\alpha\mu}B^\mu_{i\nu}+A_{q\alpha}h^{qn}(g_{\nu\mu}C^\mu_{ni}+A_{m\nu}\gamma^m_{ni0}-\partial_\nu h_{ni}-A_{n\mu}B^\mu_{i\nu})-\nabla_\alpha A_{i\nu})=0,\end{align}\begin{align}\label{BC3}
&\nabla_\beta A_{i\alpha}-\nabla_\alpha A_{i\beta}+2\imath A_{p\alpha}h^{kl}\gamma^p_{lij}(g_{\beta\mu}C^\mu_{kj}+A_{m\beta}\gamma^m_{kj0}-\partial_\nu h_{kj})\nonumber\\
&+2(g_{\alpha\mu}B^\mu_{i\beta}+A_{q\alpha}h^{qn}(g_{\beta\mu}C^\mu_{ni}+A_{m\beta}\gamma^m_{ni0}-\partial_\beta h_{ni}-A_{n\mu}B^\mu_{i\beta}))=0,\end{align}
\begin{align}\label{BC4}
&2\imath A_{p\alpha}A_{q\beta}h^{kn}h^{pl}\gamma^q_{lnm}(A_{k\mu}C^\mu_{mi}+h_{kj}\gamma^j_{mi0})-2\imath A_{p\alpha}h^{kl}\gamma^p_{lij}(g_{\beta\mu}C^\mu_{kj}+A_{m\beta}\gamma^m_{kj0}-\partial_\nu h_{kj})\nonumber\\
&+(h_{ji}-h_{ij})h^{jk}(g_{\alpha\mu}B^\mu_{k\beta}+A_{q\alpha}h^{qn}(g_{\beta\mu}C^\mu_{nk}+A_{m\beta}\gamma^m_{nk0}-\partial_\beta h_{nk}-A_{n\mu}B^\mu_{k\beta})-\nabla_\alpha A_{k\beta})=0,\end{align}\begin{align}\label{BC5}
&g_{\mu\alpha}C^\mu_{ij}+A_{m\alpha}\gamma^m_{ij0}+A_{j\mu}B^\mu_{i\alpha}+2\imath A_{p\alpha}h^{kl}\gamma^p_{lim}(A_{k\mu}C^\mu_{mj}+h_{kn}\gamma^n_{mj0})\nonumber\\
&+h_{nj}h^{nk}(g_{\alpha\mu}C^\mu_{ki}+A_{m\alpha}\gamma^m_{ki0}-\partial_\alpha h_{ki}-A_{k\mu}B^\mu_{i\alpha})=0,\end{align}\begin{align}\label{BC6}
&-\partial_\alpha h_{ji}+g_{\mu\alpha}(C^\mu_{ij}+C^\mu_{ji})+A_{m\alpha}(\gamma^m_{ij0}+\gamma^m_{ji0})+2\imath \gamma^p_{ijq}(g_{\alpha\mu}C^\mu_{pq}+A_{m\alpha}\gamma^m_{pq0}-\partial_\alpha h_{pq})=0,\end{align}
\begin{align}\label{BC7}
&A_{i\mu}C^\mu_{mn}+h_{ji}\gamma^j_{mn0}+A_{n\mu}C^\mu_{mi}+h_{nj}\gamma^j_{mi0}+2\imath \gamma^p_{mnq}(A_{p\mu}C^\mu_{qi}+h_{pj}\gamma^j_{qi0})=0,\end{align}
\begin{align}\label{BC8}
&\nabla_\beta A_{i\alpha}-\nabla_\alpha A_{i\beta}+g_{\alpha\mu}B^\mu_{i\beta}-g_{\beta\mu}B^\mu_{i\alpha}+A_{q\alpha}h^{qn}(g_{\beta\mu}C^\mu_{ni}+A_{m\beta}\gamma^m_{ni0}-\partial_\beta h_{ni}-A_{n\mu}B^\mu_{i\beta})\nonumber\\
&-A_{q\beta}h^{qn}(g_{\alpha\mu}C^\mu_{ni}+A_{m\alpha}\gamma^m_{ni0}-\partial_\alpha h_{ni}-A_{n\mu}B^\mu_{i\alpha})=0.
\end{align}
The remaining fields in $\nabla$ are determined in terms of the above by 
\begin{align}\label{DAAhh}
D^n_{\alpha\beta k}=A_{i\alpha}A_{j\beta}h^{nl}h^{im}\gamma^j_{ml k},
\end{align}
\begin{align}\label{EAh}
E^n_{j\alpha k}=A_{m\alpha}h^{ni}\gamma^m_{ij k},
\end{align}
\begin{align}\label{E0}
E^n_{j\alpha 0}=h^{ni}(g_{\alpha\mu}C^\mu_{ij}+A_{m\alpha}\gamma^m_{ij0}-\partial_\alpha h_{ij}-A_{i\mu}B^\mu_{j\alpha}),
\end{align}
\begin{align}\label{D0}
D^n_{\alpha\beta 0}=h^{ni}(g_{\alpha\mu}B^\mu_{i\beta}+A_{m\alpha}h^{mk}(g_{\beta\mu}C^\mu_{ki}+A_{m\beta}\gamma^m_{ki0}-\partial_\beta h_{ki}-A_{k\mu}B^\mu_{i\beta})-\nabla_\alpha A_{i\beta}).
\end{align}
 \end{lemma}
\proof
Substituting the form of the quantum metric into the QRG metric compatibility condition in Section~\ref{secQRG},  the first term becomes
\begin{align*}
(\nabla\tens\id)\cg&=\nabla (g_{\mu\nu}\extd x^\mu)\tens\extd x^\nu+\nabla (A_{i\mu}s^i)\tens\extd x^\mu+\nabla (A_{i\mu}\extd x^\mu)\tens s^i+\nabla (h_{ij} s^i)\tens s^j\\
&= (\extd g_{\mu\nu}\tens \extd x^\mu+g_{\mu\nu}\nabla\extd x^\mu)\tens\extd x^\nu+\extd A_{i\mu}\tens s^i\tens\extd x^\mu+A_{i\mu}\nabla s^i\tens\extd x^\mu\\
&+\extd A_{i\mu}\tens \extd x^\mu \tens s^i+A_{i\mu}\nabla\extd x^\mu\tens s^i+\extd h_{ij}\tens  s^i\tens s^j+h_{ij}\nabla s^i\tens s^j
\end{align*}
while the second term becomes
\begin{align*}
&(\sigma\tens\id)(\id\tens\nabla)\cg\\
&=(\sigma\tens\id)(g_{\mu\nu}\extd x^\mu \tens\nabla\extd x^\nu+A_{i\mu} s^i\tens \nabla\extd x^\mu+A_{i\mu}\extd x^\mu\tens\nabla s^i+h_{ij} s^i\tens\nabla s^j).
\end{align*}
After substituting $\nabla \extd x^\mu,\nabla s^i$ from Proposition~\ref{tornabla} and $\sigma$ from Lemma~\ref{lemnabla}, we find metric compatibility as the 8 equations
\begin{align*}
&\extd x^\alpha\tens\extd x^\beta\tens\extd x^\nu:\quad \partial_\alpha g_{\beta\nu}-g_{\mu\nu}\Gamma^\mu_{\alpha\beta}-g_{\beta\mu}\Gamma^\mu_{\alpha\nu}+A_{i\nu}D^i_{\alpha\beta}+A_{i\beta}D^i_{\alpha\nu}+2\imath D^i_{\alpha\beta j}(A_{i\mu}B^\mu_{j\nu}+h_{im}E^m_{j\nu})=0,\\
&\extd x^\alpha\tens s^i\tens\extd x^\beta:\quad \partial_\alpha A_{i\beta}+g_{\mu\beta}B^\mu_{i\alpha}+A_{j\beta}E^j_{i\alpha}+h_{ij}D^j_{\alpha\beta}-A_{i\mu}\Gamma^\mu_{\alpha\beta}+2\imath E^k_{i\alpha j}(A_{k\mu}B^\mu_{j\beta}+h_{km}E^m_{j\beta})=0,\\
&\extd x^\alpha\tens\extd x^\beta\tens s^i:\quad \partial_\alpha A_{i\beta}+g_{\mu\beta}B^\mu_{i\alpha}+A_{j\beta}E^j_{i\alpha}+h_{ji}D^j_{\alpha\beta}-A_{i\mu}\Gamma^\mu_{\alpha\beta}+2\imath D^k_{\alpha\beta m}(A_{k\mu}C^\mu_{mi}+h_{kj}\gamma^j_{mi})=0,\\
&\extd x^\alpha\tens s^i\tens s^j:\quad \partial_\alpha h_{ij}+A_{j\mu}B^\mu_{i\alpha}+h_{mj}E^m_{i\alpha}+A_{i\mu}B^\mu_{j\alpha}+h_{im}E^m_{j\alpha}+2\imath E^k_{i\alpha m}(A_{k\mu}C^\mu_{mj}+h_{kn}\gamma^n_{mj})=0,\\
&s^i\tens\extd x^\alpha\tens\extd x^\beta:\quad g_{\mu\beta}B^\mu_{i\alpha}+A_{j\beta}E^j_{i\alpha}+g_{\alpha\mu}B^\mu_{i\beta}+A_{j\alpha}E^j_{i\beta}+2\imath E^k_{i\alpha j}(A_{k\mu}B^\mu_{j\beta}+h_{km}E^m_{j\beta})=0,\\
&s^i\tens\extd x^\alpha\tens s^j:\quad g_{\mu\alpha}C^\mu_{ij}+A_{m\alpha}\gamma^m_{ij}+A_{j\mu}B^\mu_{i\alpha}+h_{mj}E^m_{i\alpha}+2\imath E^k_{i\alpha m}(A_{k\mu}C^\mu_{mj}+h_{kn}\gamma^n_{mj})=0,\\
&s^i\tens s^j\tens \extd x^\alpha:\quad g_{\mu\alpha}C^\mu_{ij}+A_{m\alpha}\gamma^m_{ij}+A_{j\mu}B^\mu_{i\alpha}+h_{jm}E^m_{i\alpha}+2\imath \gamma^p_{ijq}(A_{p\mu}B^\mu_{q\alpha}+h_{pm}E^m_{q\alpha})=0,\\
& s^m\tens s^n\tens s^i:\quad A_{i\mu}C^\mu_{mn}+h_{ji}\gamma^j_{mn}+A_{n\mu}C^\mu_{mi}+h_{nj}\gamma^j_{mi}+2\imath \gamma^p_{mnq}(A_{p\mu}C^\mu_{qi}+h_{pj}\gamma^j_{qi})=0,
\end{align*}
where $k,n,m,p,q\in\{1,2\}$ and all the repeated labels are summed. Taking the second equation minus the fifth equation, and the fourth equation minus the sixth equation,   we obtain
\begin{align}\label{second}
\partial_\alpha A_{i\beta}+h_{ij}D^j_{\alpha\beta}-A_{i\mu}\Gamma^\mu_{\alpha\beta}-g_{\alpha\mu}B^\mu_{i\beta}-A_{j\alpha}E^j_{i\beta}=0,
\end{align}
\begin{align}\label{fourth}
\partial_\alpha h_{ij}+A_{i\mu}B^\mu_{j\alpha}+h_{im}E^m_{j\alpha}-g_{\alpha\mu}C^\mu_{ij}-A_{m\alpha}\gamma^m_{ij}=0.
\end{align}
Hence we can replace the second and fourth equations by (\ref{second}), (\ref{fourth}). Now we can see that the $\sigma^1,\sigma^2$ components of (\ref{second}) reduce to (\ref{DAAhh}), while $\sigma^1,\sigma^2$ components of (\ref{fourth}) reduce to (\ref{EAh}). Considering $\sigma^1,\sigma^2$ components of the remaining equations, we deduce
\begin{align*}
&A_{i\nu}D^i_{\alpha\beta k}+A_{i\beta}D^i_{\alpha\nu k}+2\imath D^i_{\alpha\beta j}h_{im}E^m_{j\nu k}=0,\\
&A_{j\beta}E^j_{i\alpha k}+h_{ji}D^j_{\alpha\beta k}+2\imath D^n_{\alpha\beta m}h_{nj}\gamma^j_{mik}=0,\\
&A_{j\beta}E^j_{i\alpha k}+A_{j\alpha}E^j_{i\beta k}+2\imath E^n_{i\alpha j}h_{nm}E^m_{j\beta k}=0,\\
&A_{m\alpha}\gamma^m_{ijk}+h_{mj}E^m_{i\alpha k}+2\imath E^l_{i\alpha m}h_{ln}\gamma^n_{mjk}=0,\\
&A_{m\alpha}\gamma^m_{ijk}+h_{jm}E^m_{i\alpha k}+2\imath \gamma^p_{ijq}h_{pm}E^m_{q\alpha k}=0,\\
&h_{ji}\gamma^j_{mnk}+h_{nj}\gamma^j_{mik}+2\imath \gamma^p_{mnq}h_{pj}\gamma^j_{qik}=0.
\end{align*}
Substituting (\ref{DAAhh})(\ref{EAh}) into above six equations and considering the symmetry condition $D^k_{\alpha\beta i}=D^k_{\beta\alpha i}$, these equations reduce to the group of 7 conditions on $A, \gamma$ in the statement.

Now we analyse the $\sigma^0$ component of the eight equations by substituting (\ref{DAAhh})(\ref{EAh}). These become
\begin{align*}
&\partial_\alpha g_{\beta\nu}-g_{\mu\nu}\Gamma^\mu_{\alpha\beta}-g_{\beta\mu}\Gamma^\mu_{\alpha\nu}+A_{i\nu}D^i_{\alpha\beta 0}+A_{i\beta}D^i_{\alpha\nu 0}+2\imath A_{p\alpha}A_{q\beta}h^{ik}h^{pl}\gamma^q_{lkj}(A_{i\mu}B^\mu_{j\nu}+h_{im}E^m_{j\nu 0})=0,\\
&\partial_\alpha A_{i\beta}+h_{ij}D^j_{\alpha\beta 0}-A_{i\mu}\Gamma^\mu_{\alpha\beta}-g_{\alpha\mu}B^\mu_{i\beta}-A_{j\alpha}E^j_{i\beta 0}=0,\\
&\partial_\alpha A_{i\beta}+g_{\mu\beta}B^\mu_{i\alpha}+A_{j\beta}E^j_{i\alpha 0}+h_{ji}D^j_{\alpha\beta 0}-A_{i\mu}\Gamma^\mu_{\alpha\beta}+2\imath A_{p\alpha}A_{q\beta}h^{kn}h^{pl}\gamma^q_{lnm}(A_{k\mu}C^\mu_{mi}+h_{kj}\gamma^j_{mi0})=0,\\
&\partial_\alpha h_{ij}+A_{i\mu}B^\mu_{j\alpha}+h_{im}E^m_{j\alpha 0}-g_{\alpha\mu}C^\mu_{ij}-A_{m\alpha}\gamma^m_{ij0}=0,\\
&g_{\mu\beta}B^\mu_{i\alpha}+A_{j\beta}E^j_{i\alpha 0}+g_{\alpha\mu}B^\mu_{i\beta}+A_{j\alpha}E^j_{i\beta 0}+2\imath A_{p\alpha}h^{kl}\gamma^p_{lij}(A_{k\mu}B^\mu_{j\beta}+h_{km}E^m_{j\beta 0})=0,\\
&g_{\mu\alpha}C^\mu_{ij}+A_{m\alpha}\gamma^m_{ij0}+A_{j\mu}B^\mu_{i\alpha}+h_{mj}E^m_{i\alpha 0}+2\imath A_{p\alpha}h^{kl}\gamma^p_{lim}(A_{k\mu}C^\mu_{mj}+h_{kn}\gamma^n_{mj0})=0,\\
&g_{\mu\alpha}C^\mu_{ij}+A_{m\alpha}\gamma^m_{ij0}+A_{j\mu}B^\mu_{i\alpha}+h_{jm}E^m_{i\alpha 0}+2\imath \gamma^p_{ijq}(A_{p\mu}B^\mu_{q\alpha}+h_{pm}E^m_{q\alpha 0})=0,\\
&A_{i\mu}C^\mu_{mn}+h_{ji}\gamma^j_{mn0}+A_{n\mu}C^\mu_{mi}+h_{nj}\gamma^j_{mi0}+2\imath \gamma^p_{mnq}(A_{p\mu}C^\mu_{qi}+h_{pj}\gamma^j_{qi0})=0.
\end{align*}
From the fourth equation, we obtain (\ref{E0}). Substituting (\ref{E0}) into the second equation, we get (\ref{D0}).

Then remaining six equations reduce to (\ref{BC2}),(\ref{BC3}),(\ref{BC5})-(\ref{BC7}) and 
\begin{align*}
&\nabla_\alpha A_{i\beta}+\nabla_\beta A_{i\alpha}+2\imath A_{p\alpha}A_{q\beta}h^{kn}h^{pl}\gamma^q_{lnm}(A_{k\mu}C^\mu_{mi}+h_{kj}\gamma^j_{mi0})\\
&+(h_{ij}+h_{ji})h^{jk}(g_{\alpha\mu}B^\mu_{k\beta}+A_{q\alpha}h^{qn}(g_{\beta\mu}C^\mu_{nk}+A_{m\beta}\gamma^m_{nk0}-\partial_\beta h_{nk}-A_{n\mu}B^\mu_{k\beta})-\nabla_\alpha A_{k\beta})=0.
\end{align*}
Taking the above equation minus (\ref{BC3}), we get (\ref{BC4}). Furthermore, substituting $D^k_{\alpha\beta 0}=D^k_{\beta\alpha 0}$ into (\ref{D0}), we get condition (\ref{BC8}). \endproof

We also require reality conditions on quantum metric and our bimodule connection. 

\begin{lemma}\label{reality} (1) Reality of the quantum metric $\cg$ in Lemma~\ref{lemg}  requires 
\begin{align*}
g^*_{\mu\nu}=g_{\mu\nu},\quad A^*_{i\alpha}=A_{i\alpha},\quad h^*_{ij}=h_{ji}. 
\end{align*}
(2) The QLC in Lemma~\ref{lem_metric} is $*$-preserving if and only if 
\begin{align*}
&{\rm Im}(\Gamma^\mu_{\alpha\beta})=-C^{\mu *}_{kn}A_{i\alpha}A_{j\beta}h^{nl}h^{im}\gamma^j_{ml k},\quad \ {\rm Im}(B^\mu_{i\alpha})=C^{\mu *}_{kn}A_{m\alpha}h^{nj}\gamma^m_{ji k},\\
&{\rm Im}(C^\mu_{ij})= C^{\mu *}_{lm}\gamma^m_{ijl},\quad {\rm Im}(\gamma^k_{ij0})=\gamma^{k*}_{lm0}\gamma^m_{ijl}, \quad {\rm Im}(\gamma^k_{ijq})=\gamma^{k*}_{lmq}\gamma^m_{ijl},\\
&A_{m\alpha}({\rm Im}(h^{kj}\gamma^m_{ji q})- \gamma^{k*}_{ljq}h^{jn}\gamma^m_{ni l})=0,\\
&A_{i\alpha}A_{j\beta}({\rm Im}(h^{kl}h^{im}\gamma^j_{ml q})-\gamma^{k*}_{pnq}h^{nl}h^{im}\gamma^j_{ml p})=0,\\
&-{\rm Im}(h^{ki}\nabla_\alpha A_{i\beta})+g_{\alpha\mu}{\rm Im}(h^{ki}B^\mu_{i\beta})+A_{m\alpha}g_{\beta\mu}{\rm Im}(h^{ki}h^{mn}C^\mu_{ni})+A_{m\alpha}A_{l\beta}{\rm Im}(h^{ki}h^{mn}\gamma^l_{ni0})\\
&-A_{m\alpha}{\rm Im}(h^{ki}h^{mn}\partial_\beta h_{ni})-A_{m\alpha}A_{n\mu}{\rm Im}(h^{ki}h^{mn}B^\mu_{i\beta})=\gamma^{k*}_{ij0}A_{p\alpha}A_{n\beta}h^{jl}h^{pm}\gamma^n_{ml i},\\
&g_{\alpha\mu}{\rm Im}(h^{kj}C^\mu_{ji})+A_{m\alpha}{\rm Im}(h^{kj}\gamma^m_{ji0})-{\rm Im}(h^{kj}\partial_\alpha h_{ji})-A_{j\mu}{\rm Im}(h^{kj}B^\mu_{i\alpha})= \gamma^{k*}_{lj0}A_{m\alpha}h^{jn}\gamma^m_{ni l}.
\end{align*}
\end{lemma}
\proof For $\cg^\dagger=\cg$, this is immediate  given that $\extd x^\mu$ are classical and $s^i$ are self-adjoint. For the QLC to be $*$-preserving, we need $\nabla\circ *=\sigma\circ\dagger\circ\nabla$, which applied on the basis gives the conditions 
\begin{align*}\Gamma^{\mu}_{\alpha\beta}=\Gamma^{\mu *}_{\alpha\beta}-2\imath C^{\mu *}_{ij}D^j_{\alpha\beta i},\quad B^{\mu}_{i\alpha}=B^{\mu *}_{i\alpha}+2\imath C^{\mu *}_{kj}E^j_{i\alpha k},\quad C^\mu_{ij}=C^{\mu *}_{ij}+2\imath C^{\mu *}_{lm}\gamma^m_{ijl},\\
D^{k}_{\alpha\beta 0}=D^{k*}_{\alpha\beta 0}+2\imath \gamma^{k*}_{ij0}D^j_{\alpha\beta i},\quad E^{k}_{i\alpha 0}=E^{k*}_{i\alpha 0}+2\imath \gamma^{k*}_{lj0}E^j_{i\alpha l},\quad \gamma^k_{ij0}=\gamma^{k*}_{ij0}+2\imath\gamma^{k*}_{lm0}\gamma^m_{ijl},\\
D^{k}_{\alpha\beta q}=D^{k*}_{\alpha\beta q}+2\imath \gamma^{k*}_{ijq}D^j_{\alpha\beta i},\quad E^{k}_{i\alpha q}=E^{k*}_{i\alpha q}+2\imath \gamma^{k*}_{ljq}E^j_{i\alpha l},\quad \gamma^k_{ijq}=\gamma^{k*}_{ijq}+2\imath\gamma^{k*}_{lmq}\gamma^m_{ijl}.
\end{align*}
 As before, $i,j,k,l,m\in \{1,2\}$. We now substitute $D,E$ in terms of $A,B,C$, then the above conditions amount to the ones stated.
 \endproof

 \begin{proposition}\label{propsol+} Let $h_{ij}(x,t)s^i\tens s^j=h(x,t)(s^1\tens s^1- s^2\tens s^2)$ and let $\gamma$ be the 3-parameter QLC on $M_2(\C)$ in Section~\ref{QRGM2}.
 
 (a)  Solutions for a QRG in Lemma~\ref{lem_metric} can be classified into two cases:  
 
 (i) $A_{1\mu}=A_{2\mu}=0$ with general possibilities for $u,v,\rho$.
 
 (ii) $(v+\rho-1)A_{1\mu}=\imath (v-\rho-1)A_{2\mu}$ with $(u-1)(v-1)=-\rho^2$.
 
 We proceed with the simplest case of (ii) where
\[ u=1+ \rho,\quad v=1-\rho,\quad \rho\ne 0,\quad A_2=0\]
(there is a similar solution with $A_1=0$ and $u=1-\rho$, $v=1+\rho$).  We also assume that $A_\mu:=A_{1\mu}$ is generic and that $\tilde g_{\mu\nu}:=g_{\mu\nu}-\frac{1}{h}A_\mu A_\nu$ is invertible. 

Then all the other fields in the solution can be determined and the reality condition in Lemma~\ref{reality} solved as follows: 

(b)   $\Gamma$  is not necessarily the usual LC for usual $g_{\mu\nu}$ but characterised by solving
\[\nabla_\mu g_{\rho\nu}=\frac{1}{2h}(A_\nu \nabla_{(\rho} A_{\mu)}+A_\rho \nabla_{(\nu} A_{\mu)}).\]

(c) $h$ is a real constant and $\Gamma$ is real.

(d) The remaining parameter  $\rho(x,t)$ for the QLC on $M_2(\C)$ is imaginary but allowed to vary over spacetime. In this case $\Delta_{M_2}=0$ and $S_{M_2}={2\rho^2\over h}$ for the finite QRG. 
\end{proposition}
\proof We first consider the conditions in Lemma~\ref{lem_metric}. The fifth equation of the group of 7 conditions on $A,\gamma$ give us the two possible forms of the solution stated in (a). One can check that above these also then satisfy the other conditions in this group.

From now we only consider the $A_{1\mu}\neq 0, A_{2\mu}=0$ case as stated where $v=1-\rho,u=1+\rho,\rho\neq 0$ and we define $A_{\mu}:=A_{1\mu}$. We assume this does not identically vanish otherwise we are back in case (i). Then in Lemma~\ref{lem_metric}, (\ref{BC7}) reduces to \[C_{ij}=0.\]
And (\ref{BC6}) reduces to\[\partial_\alpha h(x,t)=0,\]
so $h$ is a constant. The remaining conditions in Lemma~\ref{lem_metric} reduce to the equation in the statement part (b) and
\begin{equation}\label{BCDsol} B^\mu{}_\nu=\frac{1}{2}\tilde{g}^{\mu\alpha}\nabla_{[\alpha} A_{\nu ]},\quad D_{\alpha\beta}=\frac{\imath}{h^2}A_{\alpha}A_{\beta},\quad D_{\alpha\beta 0}=-\frac{1}{2h}\nabla_{(\alpha}A_{\beta)},\end{equation}
\begin{equation}\label{DEsol} E_\alpha= \frac{\imath}{h}A_{\alpha},\quad E{}_{\alpha 0}=-\frac{1}{h}A_{\beta}B^\beta{}_{\alpha},
 \end{equation}
where we write 
\[ B^\mu{}_\nu =B{}_1^\mu{}_\nu,\quad D_{\alpha\beta}=D^1_{\alpha\beta1},\quad D_{\alpha\beta 0}=D^1_{\alpha\beta 0},\quad   E_\alpha=E^1_{1\alpha1},\quad E_{\alpha0}=E^1_{1\alpha0}\]
 with zero for the remaining fields (i.e. involving 2 for the latin indices). 
The non-trivial conditions in Lemma~\ref{reality} are
\[h^*=h,\quad A^*_{\alpha}=A_{\alpha},\quad (g_{\alpha\beta})^*=g_{\alpha\beta},\quad (\Gamma^\mu{}_{\alpha\beta})^*=\Gamma^\mu{}_{\alpha\beta},\]
in line with our interpretation that  $g_{\alpha\beta}$ is a usual real metric on $M$ and $\Gamma$ can be viewed as real Christoffel symbols (but not necessarily those of Levi-Civita connection for $g$). 
\endproof

This describes the simplest solution for the QRG on the product where not both $A_i$ are zero. There {\em are} more general solutions of type (ii), for example 
\begin{equation}\label{Atypeii} A_1=\pm A_2,\quad u=v=0,\quad \rho=\mp \imath\end{equation}
with all the other fields determined, $\Gamma$ the Levi-Civita connection of $g$ and $\tilde g=g$, but the scalar field action and our other results below do not have a clear physical interpretation in this case. We also saw in the proof that if $A_1=A_2=0$ then there are more possibilities for $\gamma$ and defer this case to Section~\ref{secA0}. Finally, if $\tilde g_{\mu\nu}$ in the case that we looked at is not invertible there are again more possibilities for $B^\mu{}_\nu$ and of some of the other fields. 

Sticking for the rest of the section with our simplest solution with $A_2=0$ and $A:=A_1$ in Proposition~\ref{propsol+}, the only thing we still have to solve for a QRG on the product is to find $\Gamma$ so that the condition in part (b) holds. Also note that  $h(x,t)$ in  Proposition~\ref{propsol+} is forced to be a nonzero real constant $h$. It represents the physical scale of the QRG of $M_2(\C)$. 

We first calculate the inverse $(\ ,\ )$ of the quantum metric $\cg$, which  has the general form
\begin{align}\label{tildeg}
\tilde g^{\mu\nu}=(\extd x^\mu,\extd x^\nu),\quad \tilde A^{\mu i}=(\extd x^\mu,s^i),\quad \tilde A^{i\mu}=(s^i,\extd x^\mu),\quad \tilde h^{ij}=(s^i,s^j)
\end{align}
but we will show that  $\tilde g^{\mu\nu}$ is indeed the inverse of $\tilde g_{\mu\nu}$ as the notation implies. Using 
the definition of the inverse quantum metric in Section~\ref{secQRG}, we find that the above coefficients are required to obey
 \begin{align*}
 A_{i\mu}\tilde A^{\alpha i}+g_{\mu\nu}\tilde g^{\alpha\nu}=\delta^\alpha_\mu,\quad A_{i\mu}\tilde A^{i\alpha}+g_{\mu\nu}\tilde g^{\nu\alpha}=\delta^\alpha_\mu,\quad A_{i\mu}\tilde A^{k\mu}+h_{ji}\tilde h^{kj}=\delta^k_i,\quad A_{i\mu}\tilde A^{\mu k}+h_{ij}\tilde h^{jk}=\delta^k_i,\\
 A_{i\mu}\tilde g^{\alpha\mu}+h_{ji}\tilde A^{\alpha j}=0,\quad A_{i\mu}\tilde g^{\mu\alpha}+h_{ij}\tilde A^{j\alpha}=0,\quad A_{i\mu}\tilde h^{ki}+g_{\mu\nu}\tilde A^{k\nu}=0,\quad A_{i\mu}\tilde h^{ik}+g_{\mu\nu}\tilde A^{\nu k}=0,
\end{align*}  
which can  be reduced in our case to 
\begin{align}\label{tilde}
\tilde g^{\mu\nu}=g^{\mu\nu}+\frac{g^{\mu\alpha}g^{\nu\beta}A_{\alpha}A_{\beta}}{h- g^{\rho\sigma}A_{\rho}A_{\sigma}},\quad\tilde A^{\mu}:=-\frac{1}{h}\tilde g^{\mu\nu}A_{\nu}=-\frac{g^{\mu\nu}A_{\nu}}{h- g^{\mu\nu}A_{\mu}A_{\nu}},\\
\tilde h^{11}=\frac{1}{h- g^{\mu\nu}A_{\mu}A_{\nu}},\quad\tilde h^{22}=-\frac{1}{h},\quad \tilde A^{\alpha 2}=\tilde A^{2\alpha}=\tilde h^{12}=\tilde h^{21}=0,
\end{align}
where we denote $\tilde A^{\mu}=\tilde A^{1\mu}$. The first of these is equivalent to  $\tilde g^{\mu\nu}$ being the inverse of  $\tilde g_{\mu\nu}=g_{\mu\nu}-\frac{1}{h}A_\mu A_\nu$ as in Proposition~\ref{propsol+}.

We also define
 $\tilde\nabla \extd x^\mu=-\tilde \Gamma^\mu_{\alpha\beta}\extd x^\alpha\tens \extd x^\beta$ for the Levi-Civita connection and Christoffel symbols for $\tilde g^{\mu\nu}$ regarded as a second metric on the classical manifold.

 \begin{theorem}\label{thmmain}
Generic solutions of the equation in Proposition~\ref{propsol+}(b) and hence of QRGs on the product are provided by $A_\mu$ any 1-form and 
\[ \Gamma^\sigma_{\mu\nu}=\tilde\Gamma^\sigma_{\mu\nu}+\frac{1}{2h}\tilde g^{\sigma\rho}(A_\mu F_{\nu\rho}+A_\nu F_{\mu\rho})\]
 where $\tilde g$ is defined as in (\ref{tilde}), $\tilde \Gamma$ is its classical Levi-Civita connection and $F_{\mu\nu}=\del_\mu A_\nu-\del_\nu A_\mu$. Moreover,
\begin{align}\label{lam3}
\Gamma^\sigma_{\mu\nu}=\Lambda^\sigma_{\mu\nu}-\frac{1}{2h} g^{\sigma\rho}A_\rho \nabla_{(\mu}A_{\nu)}
\end{align}
in terms of Levi-Civita connection $\Lambda$ for our original metric $g$. 
\end{theorem}
\proof
Using the classical formula for $\Lambda$, we first show that $\Lambda$ can be written as
\begin{align}
\Lambda^\sigma_{\mu\nu}&=\frac{1}{2} g^{\sigma\rho}\left( g_{\rho\mu,\nu}+ g_{\nu\rho,\mu}- g_{\mu\nu,\rho}\right)\label{lam1}\\
&=\frac{1}{2} g^{\sigma\rho}\left( \nabla_\nu g_{\rho\mu}+ \nabla_\mu g_{\nu\rho}- \nabla_\rho g_{\mu\nu}+2\Gamma^\lambda_{\mu\nu}g_{\lambda \rho}\right)\nonumber\\
&=\Gamma^\sigma_{\mu\nu}+\frac{1}{2} g^{\sigma\rho}\left( \nabla_\nu g_{\rho\mu}+ \nabla_\mu g_{\nu\rho}- \nabla_\rho g_{\mu\nu}\right)\label{lam2}
\end{align}
where $\nabla_\mu$ is defined by $\Gamma$ in our coordinates. Then we compute
\begin{align*}
&\tilde\Gamma^\sigma_{\mu\nu}=\frac{1}{2}\tilde g^{\sigma\rho}(\tilde g_{\rho\mu,\nu}+\tilde g_{\nu\rho,\mu}-\tilde g_{\mu\nu,\rho})\\
&=\frac{1}{2}\left( g^{\sigma\rho}+\frac{g^{\sigma\alpha}g^{\rho\beta}A_\alpha A_\beta}{h-g^{\lambda\gamma}A_\lambda A_\gamma}\right)\left( g_{\rho\mu,\nu}+ g_{\nu\rho,\mu}- g_{\mu\nu,\rho}-\frac{1}{h}(\del_{[\nu}A_{\rho]}A_\mu+\del_{(\nu}A_{\mu)}A_\rho+\del_{[\mu}A_{\rho]}A_\nu)\right)\\
&=\Lambda^\sigma_{\mu\nu}+\frac{g^{\sigma\alpha}\Lambda^\beta_{\mu\nu}A_\alpha A_\beta}{h-g^{\lambda\gamma}A_\lambda A_\gamma}-\frac{1}{2h} \tilde g^{\sigma\rho}\left(\nabla_{[\nu}A_{\rho]}A_\mu+\nabla_{(\nu}A_{\mu)}A_\rho+\Gamma^\lambda_{(\nu\mu)}A_\lambda A_\rho+\nabla_{[\mu}A_{\rho]}A_\nu\right)\\
&=\Gamma^\sigma_{\mu\nu}+\frac{1}{2} g^{\sigma\rho}\left( \nabla_\nu g_{\rho\mu}+ \nabla_\mu g_{\nu\rho}- \nabla_\rho g_{\mu\nu}\right)+\frac{g^{\sigma\alpha}\Gamma^\beta_{\mu\nu}A_\alpha A_\beta}{h-g^{\lambda\gamma}A_\lambda A_\gamma}\\
&\quad+\frac{1}{2} g^{\beta\rho}\left( \nabla_\nu g_{\rho\mu}+ \nabla_\mu g_{\nu\rho}- \nabla_\rho g_{\mu\nu}\right)\frac{g^{\sigma\alpha}A_\alpha A_\beta}{h-g^{\lambda\gamma}A_\lambda A_\gamma}\\
&\quad-\frac{1}{2h} \tilde g^{\sigma\rho}\left(\nabla_{[\nu}A_{\rho]}A_\mu+\nabla_{(\nu}A_{\mu)}A_\rho+\Gamma^\lambda_{(\nu\mu)}A_\lambda A_\rho+\nabla_{[\mu}A_{\rho]}A_\nu\right)\\
&=\Gamma^\sigma_{\mu\nu}+\frac{1}{2} \tilde g^{\sigma\rho}\left( \nabla_\nu g_{\rho\mu}+ \nabla_\mu g_{\nu\rho}- \nabla_\rho g_{\mu\nu}\right)+\left(-\frac{1}{2h} \tilde g^{\sigma\rho}\Gamma^\lambda_{(\nu\mu)}A_\lambda A_\rho+\frac{g^{\sigma\alpha}\Gamma^\beta_{\mu\nu}A_\alpha A_\beta}{h-g^{\lambda\gamma}A_\lambda A_\gamma}\right)\\
&\quad-\frac{1}{2h} \tilde g^{\sigma\rho}\left(\nabla_{[\nu}A_{\rho]}A_\mu+\nabla_{(\nu}A_{\mu)}A_\rho+\nabla_{[\mu}A_{\rho]}A_\nu\right)\\
&=\Gamma^\sigma_{\mu\nu}+\frac{1}{2} \tilde g^{\sigma\rho}\left( \nabla_\nu g_{\rho\mu}+ \nabla_\mu g_{\nu\rho}- \nabla_\rho g_{\mu\nu}\right)-\frac{1}{2h} \tilde g^{\sigma\rho}\left(\nabla_{[\nu}A_{\rho]}A_\mu+\nabla_{(\nu}A_{\mu)}A_\rho+\nabla_{[\mu}A_{\rho]}A_\nu\right)
\end{align*}
where we use the classical formula for $\tilde\Gamma$  in  first step, formula (\ref{tilde}) for $\tilde g$ in second step, (\ref{lam1}) in the third step, (\ref{lam2}) in fourth step. The expression in the big brackets after the fifth step vanishes after substituting $\tilde g$ in terms of $g$.

Next, we define $\Gamma$ in terms of $\tilde\Gamma$ as stated in the theorem and inserting into the above calculation, 
we find
\begin{align}\label{3gAA}
 \nabla_\nu g_{\rho\mu}+ \nabla_\mu g_{\nu\rho}- \nabla_\rho g_{\mu\nu}= {1\over h} A_\rho\nabla_{(\nu}A_{\mu)}.
\end{align}
By adding the above equation and the same equation where $\rho,\mu$ are swapped, we obtain the equation in Proposition~\ref{propsol+}(b). Finally, (\ref{lam3}) can be obtained by substituting (\ref{3gAA}) into (\ref{lam2}).
\endproof

This theorem is our main result and tells us that we can solve for the QRG on the product for any initial classical metric $g$ on $M$, the given QRG on $M_2(\C)$  with constant scale factor $h$, and any generic 1-form  $A_\mu$ on $M$.  We now turn to the physical interpretation. 

\begin{lemma}\label{3eq} For $\tilde g$ defined by $A_\mu$ as above, we have: 

 (1)  $\tilde g^{\alpha\beta}\tilde\nabla_\alpha A_\beta=\tilde g^{\alpha\beta}\nabla_\alpha A_\beta$;
 
 (2) Let  $||A||^2:=\tilde g^{\alpha\beta}A_\alpha A_\beta$ then $g^{\alpha\beta}A_\alpha A_\beta=\frac{ h ||A||^2}{h+||A||^2}$;
 
 (3) If $A_\mu$ is Killing with respect to $\tilde \nabla$ then  $\tilde g^{\alpha\beta}A_\alpha F_{\beta\mu}=-\del_\mu ||A||^2$.
\end{lemma}
\proof 
For (1), we show that
\begin{align*}
\tilde g^{\alpha\beta}\tilde\nabla_\alpha A_\beta&=\tilde g^{\alpha\beta}(\del_\alpha A_\beta-\tilde \Gamma^\mu_{\alpha\beta}A_\mu)\\
&=\tilde g^{\alpha\beta}\left(\del_\alpha A_\beta-\Gamma^\mu_{\alpha\beta}A_\mu+\frac{1}{2h}\tilde g^{\mu\nu}(A_\beta F_{\alpha\nu}+A_\alpha F_{\beta\nu})A_\mu\right)\\
&=\tilde g^{\alpha\beta}(\del_\alpha A_\beta-\Gamma^\mu_{\alpha\beta}A_\mu)\\
&=\tilde g^{\alpha\beta}\nabla_\alpha A_\beta.
\end{align*}
For (2), we deduce
\begin{align*}
||A||^2&=\tilde g^{\alpha\beta}A_\alpha A_\beta=\left(g^{\alpha\beta}+\frac{g^{\alpha\mu}g^{\beta\nu}A_\mu A_\nu}{h-g^{\rho\sigma}A_\rho A_\sigma}\right)A_\alpha A_\beta=g^{\alpha\beta}A_\alpha A_\beta+\frac{(g^{\alpha\mu}A_\alpha A_\mu)(g^{\beta\nu}A_\beta A_\nu)}{h-g^{\rho\sigma}A_\rho A_\sigma}\\
&=\frac{ h g^{\alpha\beta}A_\alpha A_\beta}{h-g^{\rho\sigma}A_\rho A_\sigma}
\end{align*}
from which $g^{\alpha\beta}A_\alpha A_\beta$ can be written as stated. For (3), 
\[ \del_\mu ||A||^2= 2\tilde g^{\alpha\beta}A_\alpha \tilde\nabla_{\mu} A_{\beta}=\tilde g^{\alpha\beta}A_\alpha \tilde\nabla_{(\mu} A_{\beta)} - \tilde g^{\alpha\beta}A_\alpha F_{\beta\mu}\]
from which the statement follows. \endproof

We also note that the classical Laplacian for a real scalar field with respect to our second metric can be computed as 
\begin{align*}
&\tilde\Delta_{LB} = \tilde g^{\alpha\beta}(\partial_\alpha\partial_\beta -\tilde\Gamma^\mu_{\alpha\beta}\partial_\mu )=\tilde g^{\alpha\beta}
(\partial_\alpha\partial_\beta -(\Gamma^\mu_{\alpha\beta}-\frac{1}{h}\tilde g^{\mu\nu}A_\beta F_{\alpha\nu})\partial_\mu)\end{align*}
in view of Theorem~\ref{thmmain}.

Now working in the background of the solution in Theorem~\ref{thmmain}, we work out the quantum geometric Laplacian $\Delta$ on a scalar field $f=\sum f_a(x,t)\sigma^a$ (where $a=0,1,2,3)$ and write down the associated action. 

\begin{proposition}\label{proplap} The action for a massless real scalar field $f$ on the product with $f_2,f_3$ viewed as a single complex scalar field $\psi = f_2 - \imath f_3$  is
\[ S_f=-\int_M\sqrt{-\tilde g}\prod\extd x^\mu \left(\sum_{a=0,1} (\tilde\nabla_\mu f_a) \tilde\nabla^\mu f_a+ \overline{(\tilde\nabla_{\mu A}\psi)}\tilde\nabla^\mu_{A}\psi\right)\]
where 
\[ \tilde\nabla_{\mu A}:=\tilde \nabla_\mu - \imath \frac{A_\mu}{h}.\]\end{proposition}
\proof  We start by computing the QRG Laplacian for the real scalar field $f$ in terms of its components,
\begin{align*}
&\Delta f=(\ , )\nabla \extd (f_a\sigma^a)\\
=&((\partial_\alpha\partial_\beta f_a-(\Gamma^\mu_{\alpha\beta}-\frac{2}{h}\tilde g^{\mu\nu}A_{\beta}\nabla_\alpha A_\nu)\partial_\mu f_a)\sigma^a-\frac{1}{h^2}A_{\alpha}A_{\beta}(f_{2}\sigma^{2}+f_3\sigma^3)\\
&-\frac{2}{h}A_{\beta}((\partial_\alpha f_3) \sigma^{2}-(\partial_\alpha f_{2}) \sigma^3))\tilde g^{\alpha\beta}-\frac{1}{h}(f_3\sigma^2-f_2\sigma^3)\tilde g^{\alpha\beta}\tilde\nabla_\alpha A_\beta \\
=&\tilde\Delta_{LB}f-\frac{1}{h^2}\tilde g^{\alpha\beta}A_{\alpha}A_{\beta}(f_{2}\sigma^{2}+f_3\sigma^3)-\frac{2}{h}\tilde g^{\alpha\beta}A_{\beta}\left(\sigma^{2}\partial_\alpha f_3 -\sigma^3\partial_\alpha f_{2}\right)-\frac{1}{h}(f_3\sigma^2-f_2\sigma^3)\tilde g^{\alpha\beta}\tilde\nabla_\alpha A_\beta
\end{align*}
where $\tilde\Delta_{LB}$ acts on each component  $f_a$ when $f$ is regarded as a multiplet on spacetime.

Next, the action for the massless real scalar field on the tensor product spacetime is given by integration of $f \Delta f=\sum_{a,b}f_a(\Delta f)_b\sigma^a\sigma^b$ over $M$ and over $M_2(\C)$ where, by the latter, we mean the normalised trace ${1\over 2}{\rm Tr}$ as a positive linear functional $M_2(\C)\to \C$. In terms of the fields $f_a$ viewed as a multiplet on
$M$, this becomes
\begin{align*}
&S_f= \int_M\sqrt{-\tilde g} \prod\extd x^\mu\sum_jf_a (\Delta f)_a\\
&=\int_M\sqrt{-\tilde g}\prod\extd x^\mu \left(\sum_a  f_a\tilde\Delta_{LB} f_a-\frac{1}{h^2}\tilde g^{\alpha\beta}A_\alpha A_\beta ( f_{2}f_{2}+  f_3 f_3)-\frac{2}{h}\tilde g^{\alpha\beta}A_\beta( f_{2}\partial_\alpha  f_3 -f_3\partial_\alpha  f_{2})\right).
\end{align*}
Finally, we change variables to $\psi$ as stated to give
\[ S_f=\int_M\sqrt{-\tilde g}\prod\extd x^\mu \left(\sum_{a=0,1} f_a\tilde\Delta_{LB} f_a+ {1\over 2}(\bar\psi\tilde\Delta_{LB,A}\psi+h.c.)\right)\] 
for $\tilde \Delta_{LB,A}=\tilde g^{\mu\nu}\tilde\nabla_{\mu A}\tilde\nabla_{\nu A}$, which we then write as stated. \endproof

This exhibits a real massless scalar field $f$ on the total space as a multiplet of massless fields on $M$ where $f_0, f_1$ have zero charge and $f_2,f_3$ combine to a charged scalar field with $A_\mu$ entering like a gauge field.

\begin{theorem}\label{thmS} For our  $A_\mu$ solutions, 
\[S=\tilde S_M + {2 \rho^2 \over h} +\frac{1}{8h}|| F||^2+\frac{1}{8h^2} ||A.F||^2,\]
where $A.F_\mu=\tilde g^{\alpha\beta} A_\alpha F_{\beta\mu}$, $||\ ||^2$ likewise contracts with $\tilde g$, $F=\extd A$ as in Lemma~\ref{3eq} and $\tilde S_M$ is the Ricci scalar of $\tilde g$ in our conventions. 
\end{theorem}
\proof From the general form of $\nabla$ and the some of the properties
of the fields $B,D, E,\Gamma$ in the proof of Proposition~\ref{propsol+}, we find
\begin{align*}
R_\nabla\extd x^\mu&=(-\partial_\nu \Gamma^\mu_{\alpha\beta}-\Gamma^\mu_{\nu\rho}\Gamma^\rho_{\alpha\beta}-B^\mu_{\ \nu}D^1_{\alpha\beta})\extd x^\nu\wedge\extd x^\alpha\tens\extd x^\beta\\
&+(\partial_\alpha B^\mu_{\ \beta}+\Gamma^\mu_{\alpha\rho}B^\rho_{\ \beta}-B^\mu_{\ \alpha}E^1_{1\beta})\extd x^\alpha\wedge\extd x^\beta\tens s^1\\
&+(\partial_\alpha B^\mu_{\ \beta}+\Gamma^\mu_{\alpha\rho}B^\rho_{\ \beta}-B^\mu_{\ \rho}\Gamma^\rho_{\alpha\beta}-B^\mu_{\ \alpha}E^1_{1\beta})\extd x^\alpha\wedge s^1\tens\extd x^\beta\\
&+(B^\mu_{\ \rho}B^\rho_{\ \alpha}-\imath B^\mu_{\ \alpha}\sigma^1)\extd x^\alpha\wedge s^1\tens s^1-(\imath B^\mu_{\ \rho}B^\rho_{\ \alpha}+B^\mu_{\ \alpha}\sigma^1){\rm Vol}\tens \extd x^\alpha,
\end{align*}
\begin{align*}
R_\nabla s^1&=(\partial_\nu D^1_{\alpha\beta}+D^1_{\nu\rho}\Gamma^\rho_{\alpha\beta}-E^1_{1\nu}D^1_{\alpha\beta})\extd x^\nu\wedge\extd x^\alpha\tens\extd x^\beta\\
&+(\partial_\alpha E^1_{1\beta}-D^1_{\alpha\rho}B^\rho_{\ \beta}-E^1_{1\alpha}E^1_{1\beta})\extd x^\alpha\wedge\extd x^\beta\tens s^1\\
&+(D^1_{\alpha\beta }\epsilon_{i13}\sigma^3+\partial_\alpha E^1_{i\beta}-D^1_{\alpha\rho}B^\rho_{i\beta}-E^1_{i\rho}\Gamma^\rho_{\alpha\beta}-E^1_{1\alpha}E^1_{i\beta}+\gamma^1_{i11}\sigma^1 D^1_{\alpha\beta})\extd x^\alpha\wedge s^i\tens\extd x^\beta\\
&+(E_{\alpha}\epsilon_{i13}\sigma^3+E^1_{i\rho}B^\rho_{\ \alpha})\extd x^\alpha\wedge s^i\tens s^1-\imath E^1_{1\rho}B^\rho_{\ \alpha}{\rm Vol}\tens \extd x^\alpha,
\end{align*}
\begin{align*}
R_\nabla s^2&=-4\imath\rho^2{\rm Vol}\tens s^2-2(\del_\mu \rho) \sigma^2\extd x^\mu\wedge s^1\tens s^2.
\end{align*}
This gives the Ricci tensor and Ricci scalar
\begin{align*}
{\rm Ricci}&={1\over 2}((-\partial_\mu \Gamma^\mu_{\alpha\beta}-\Gamma^\mu_{\mu\rho}\Gamma^\rho_{\alpha\beta}+\partial_\alpha \Gamma^\mu_{\mu\beta}+\Gamma^\mu_{\alpha\rho}\Gamma^\rho_{\mu\beta}-\nabla_\alpha E^1_{1\beta}-2\imath D^1_{\alpha\beta}\sigma^1\\
&+E^1_{1\alpha}E^1_{1\beta}+D^1_{\mu(\alpha}B^\mu_{\ \beta)})\extd x^\alpha\tens\extd x^\beta+(\nabla_\mu B^\mu_{\ \alpha}-\imath \sigma^1 E^1_{1\alpha})\extd x^\alpha\tens s^1\\
&+(\nabla_\mu B^\mu_{\ \alpha}- E^1_{1\rho}B^\rho_{\ \alpha}) s^1\tens\extd x^\alpha+B^\mu_{\ \rho}B^\rho_{\ \mu} s^1\tens s^1)-2 \rho^2 s^2\tens s^2,
\end{align*}
\begin{align*}
S&={1\over 2}\left(-\partial_\mu \Gamma^\mu_{\alpha\beta}+\partial_\alpha \Gamma^\mu_{\mu\beta}+\Gamma^\mu_{\alpha\rho}\Gamma^\rho_{\mu\beta}-\Gamma^\mu_{\mu\rho}\Gamma^\rho_{\alpha\beta}+\frac{1}{h}\nabla_\alpha (A_\mu B^\mu_\beta)+\frac{1}{h^2}A_\mu A_\nu B^\mu_\alpha B^\nu_\beta\right.\\
&-\left.\frac{2}{h}A_\alpha\nabla_\mu B^\mu_{\ \beta}- \frac{1}{h^2}A_\nu B^\nu_\rho B^\rho_{\ \alpha}A_\beta\right)\tilde g^{\alpha\beta}+\frac{1}{2}B^\mu_{\ \rho}B^\rho_{\ \mu} \tilde h^{11}+\frac{2 \rho^2}{h}. 
\end{align*}
Next, we recognise part of $S$ as the Ricci scalar for $\tilde g$, which using $\tilde\Gamma$ from Proposition~\ref{thmmain}, is
\begin{align*}
\tilde S_M&={1\over 2}\left( -\partial_\mu\tilde\Gamma^\mu_{\alpha\beta}+\partial_\alpha \tilde\Gamma^\mu_{\mu\beta}+\tilde\Gamma^\mu_{\alpha\rho}\tilde\Gamma^\rho_{\mu\beta}-\tilde\Gamma^\mu_{\mu\rho}\tilde\Gamma^\rho_{\alpha\beta}\right)\\
&={1\over 2}\left( -\partial_\mu\Gamma^\mu_{\alpha\beta}+\partial_\alpha \Gamma^\mu_{\mu\beta}+\Gamma^\mu_{\alpha\rho}\Gamma^\rho_{\mu\beta}-\Gamma^\mu_{\mu\rho}\Gamma^\rho_{\alpha\beta}+\frac{1}{h}\nabla_\alpha (A_\mu B^\mu_{\ \beta})-\frac{2}{h}\nabla_\mu(A_\alpha B^\mu_{\ \beta})\right.\\
&\quad+\left.\frac{1}{h^2}(A_\mu A_\nu B^\mu_{\ \alpha} B^\nu_{\ \beta}+A_\alpha B^\mu_{\ \rho}A_\beta B^\rho_{\ \mu})\right)\tilde g^{\alpha\beta}
\end{align*}
so that 
\begin{align*}
S&=\tilde S_M + {2 \rho^2 \over h}+\frac{1}{2}\left(- \frac{1}{h^2}A_\nu B^\nu_\rho B^\rho_{\ \alpha}A_\beta+\frac{2}{h} B^\mu_{\ \beta}\nabla_\mu A_\alpha-\frac{1}{h^2}A_\alpha B^\mu_{\ \rho}A_\beta B^\rho_{\ \mu}\right)\tilde g^{\alpha\beta}+\frac{1}{2}B^\mu_{\ \rho}B^\rho_{\ \mu} \tilde h^{11}\\
&=\tilde S_M + {2 \rho^2 \over h}+\frac{1}{2}\left(-\frac{2}{h} -\frac{1}{h^2}\tilde g^{\alpha\beta}A_\alpha A_\beta+ \tilde h^{11} \right)B^\mu_{\ \rho}B^\rho_{\ \mu}+\frac{1}{2h}\left(-\frac{1}{ h}A_\nu B^\nu_{\ \rho} B^\rho_{\ \alpha}A_\beta+B^\mu_{\ \beta}\nabla_{(\alpha} A_{\mu)}\right)\tilde g^{\alpha\beta}\\
&=\tilde S_M + {2 \rho^2 \over h} -\frac{1}{2h}\left(B^\mu_{\ \rho} B^\rho_{\ \mu}+\frac{1}{h}A_\nu B^\nu_{\ \rho} B^\rho_{\ \alpha}A_\beta \tilde g^{\alpha\beta}\right). 
\end{align*}
Finally, we write  $B^\mu{}_\nu$ in terms of $\nabla_{[\mu}A_{\nu]}=F_{\mu\nu}$ according to (\ref{BCDsol}), to  give the result for $S$ as stated.  \endproof

The $\rho$ here is in principle allowed to vary on spacetime, but this does not appear to be associated to interesting dynamics so we will take it a constant (i.e. fix both the quantum metric and connection in $M_2(\C)$), in which case this term can be dropped as a constant from the action. The $||A.F||^2$ term is unexpected but Lemma~\ref{3eq} says (by way of getting some intuition) that if we make a simplifying `gauge fixing' like  assumption that $A_\mu$ is a Killing form with respect to $\tilde \nabla$ then this term is  
\[ ||A.F||^2=\tilde g^{\mu\nu}\del_\mu(||A||^2)\del_\nu(||A||^2)\]
which is  like a kinetic term for $||A||^2$ as a real scalar field. 

\section{Special case where $A=0$}\label{secA0}

In this section, we impose a second simplifying condition on $\sigma$ in addition to (\ref{condi}), namely that $\sigma(s^i\tens s^j)$ doesn't have any terms with $\extd x^\mu$. According to (\ref{ssij}), this means
\begin{align}\label{DE}
D^i_{\alpha\beta j}=E^i_{k\alpha j}=0.
\end{align}
With this further condition, we can solve more general $h_{ij}(x,t)$ but with $A=0$, in contrast to the preceding section. This is the same as case (i) in Proposition~\ref{propsol+}. 

\begin{proposition}\label{propA0} Assuming (\ref{DE}),  a QLC of the form in Proposition~\ref{tornabla} has $A=0$, $\Gamma$ the usual Levi-Civita connection for  $g_{\mu\nu}$, $\gamma$ a QLC for $h_{ij}$ on $M_2(\C)$ according to
\begin{align}\label{hg}
 h_{ji}\gamma^j_{mn}+2\imath h_{pj}\gamma^p_{mnq}\gamma^j_{qi}+h_{nj}\gamma^j_{mi}=0
\end{align}
and the fields $C,E$ required to obey
\begin{align}
&C^\mu_{ij}=- g^{\mu\alpha}h_{mj} E^m_{i\alpha 0},\label{ce}\\
&\partial_\alpha h_{ij}+h_{mj}E^m_{i\alpha 0}+h_{im}E^m_{j\alpha 0}=0,\label{he}\\
&2\imath h_{pm}E^m_{n\alpha 0}\gamma^p_{ijn}+(h_{jm}-h_{mj})E^m_{i\alpha 0}=0.\label{heg}
\end{align}
Here, $i,j,n,m,p,q\in\{1,2\}$, all the repeated labels are summed, and all fields could depend on spacetime.
\end{proposition}
\proof We use the analysis in Lemma~\ref{lem_metric} with the further assumption (\ref{DE}). Using $E^i_{k\alpha j}=0$, $E^n_{j\alpha k}=A_{m\alpha}h^{ni}\gamma^m_{ij k}$ and $\gamma^m_{11}+\gamma^m_{22}=i\sigma^m$, we find $A_{m\alpha}\imath \sigma^m=0$, which requires \[A_{m\alpha}=0\]
for all $m,\alpha$. Then the various conditions on $A$ in Lemma~\ref{lem_metric} are satisfied automatically. The conditions on $B,C$ reduce to
\begin{align*}
&\nabla_\alpha g_{\beta\nu}=0,\quad B^\mu_{i\alpha}=0,\quad C^\mu_{11}+C^\mu_{22}=0,\\
&g_{\mu\alpha}C^\mu_{ij}+h_{nj}h^{nk}(g_{\alpha\mu}C^\mu_{ki}-\partial_\alpha h_{ki})=0,\\
&-\partial_\alpha h_{ji}+g_{\mu\alpha}(C^\mu_{ij}+C^\mu_{ji})+2\imath \gamma^p_{ijq}(g_{\alpha\mu}C^\mu_{pq}-\partial_\alpha h_{pq})=0,\\
&h_{ji}\gamma^j_{mn0}+h_{nj}\gamma^j_{mi0}+2\imath \gamma^p_{mnq}h_{pj}\gamma^j_{qi0}=0.
\end{align*}
Considering the last equation of above equations and the condition on $\gamma$ in Lemma~\ref{lem_metric}, one obtains the complete condition on $\gamma$ as stated in this proposition. In addition $D^n_{\alpha\beta 0},E^n_{j\alpha 0}$ reduce to
\begin{align*}
&D^n_{\alpha\beta 0}=0,\quad E^n_{j\alpha 0}=h^{ni}(g_{\alpha\mu}C^\mu_{ij}-\partial_\alpha h_{ij}).
\end{align*}
Considering above equation about $E^n_{j\alpha 0}$, the conditions on $C$ can be written as the one stated.
 \endproof

The inverse of the quantum metric $(,)$ on the tensor product algebra, has the form
\begin{align}
(\extd x^\mu,\extd x^\nu)=g^{\mu\nu},\quad (s^i,s^j)=h^{ij},\quad (\extd x^\mu,s^i)=0,\quad (s^i,\extd x^\mu)=0
\end{align}
where $g^{\mu\nu},h^{ij}$ are the inverse of $g_{\mu\nu},h_{ij}$ respectively. It follows that  $h^{11}+h^{22}=0$ and in fact that 
\[ h^{ij}=-h_{ij}/\det(h)\]
 in terms of $h=(h_{ij})$ and its determinant.  We are now ready to compute the QRG Laplacian $\Delta=(\ ,\ )\nabla\extd$ on the tensor product, i.e. acting on any $f=f_a(x,t)\sigma^a\in C^\infty(M)\tens M_2(\C)$ using our extended Pauli basis with $\sigma^0=\id$. 
 
\begin{theorem}\label{thmA0} The QRG Laplacian on $f=f_a(x,t)\sigma^a$ in the background solution in Proposition~\ref{propA0} is
\[ \Delta f=\Delta_{LB} f  + \Delta_{M_2} f-g^{\mu\alpha}h_{mj}E^m_{i\alpha 0}(\partial_\mu f_a)\sigma^a h^{ij}\]
where $\Delta_{LB}$ is the classical Laplacian for the metric $g$ on each component $f_a$ and 
\[ \Delta_{M_2} f=h^{ij} (f_i\sigma^j-\epsilon_{kcb} f_c\sigma^{b}\gamma^k_{ij})\]
is the Laplacian on $M_2(\C)$ at each $x,t$. Here $i,j,k\in \{1,2\}, b,c\in \{1,2,3\}$ and $a\in\{0,1,2,3\}$. 
\end{theorem}
\proof We compute 
 \begin{align*}
\Delta f&=(\ , )\nabla \extd (f_a\sigma^a)=(\ , )\nabla ((\extd f_a)\sigma^a+f_a\extd\sigma^a)=(\ , )\nabla ((\partial_\mu f_a)\sigma^a \extd x^\mu +f_a(\partial_i\sigma^a) s^i)\\
&=(\ , )(\extd ((\partial_\mu f_a)\sigma^a)\tens \extd x^\mu+(\partial_\mu f_a)\sigma^a\nabla\extd x^\mu+\extd (f_a(\partial_i\sigma^a))\tens s^i+f_a(\partial_i\sigma^a)\nabla s^i)\\
&=(\ , )(((\extd (\partial_\mu f_a))\sigma^a+(\partial_\mu f_a)\extd \sigma^a)\tens \extd x^\mu+(\partial_\mu f_a)\sigma^a\nabla\extd x^\mu\\
&\quad+((\extd f_a)(\partial_i\sigma^a)+f_a \extd (\partial_i\sigma^a))\tens s^i+f_a(\partial_i\sigma^a)\nabla s^i).
\end{align*}
Substituting $\nabla \extd x^\mu,\nabla s^1,\nabla s^2$, the above result becomes
\begin{align*}
\Delta f&=(\partial_\nu\partial_\mu f_a)\sigma^a g^{\nu\mu}+(\partial_\rho f_a)\sigma^a(-\Gamma^\rho_{\mu\nu}g^{\mu\nu}+C^\rho_{ij}h^{ij})+f_a (\partial_j\partial_i\sigma^a)h^{ji}+f_a(\partial_k\sigma^a)\gamma^k_{ij}h^{ij}\\
&=g^{\mu\nu}(\partial_\mu\partial_\nu f_a-\Gamma^\rho_{\mu\nu}\partial_\rho f_a)\sigma^a+(f_a (\partial_i\partial_j\sigma^a+(\partial_k\sigma^a)\gamma^k_{ij})+C^\mu_{ij}(\partial_\mu f_a)\sigma^a)h^{ij}\\
&=(\Delta_{LB} f_a)\sigma^a+f_c( \partial_i\partial_j\sigma^c+(\partial_k\sigma^c)\gamma^k_{ij})h^{ij}+C^\mu_{ij}(\partial_\mu f_a)\sigma^a h^{ij}\\
&=(\Delta_{LB} f_a)\sigma^a+f_c( \partial_i\partial_j\sigma^c+(\partial_k\sigma^c)\gamma^k_{ij})h^{ij}-g^{\mu\alpha}h_{mj}E^m_{i\alpha 0}(\partial_\mu f_a)\sigma^a h^{ij},
\end{align*}
where $\Delta_{LB}$ is the usual  Laplacian on spacetime, $\del_i\sigma^b=-\epsilon_{ibc}\sigma^c$ and $i,j,k\in \{1,2\},a\in\{0,1,2,3\}$, $b,c\in \{1,2,3\}$ and $\epsilon_{kbc}$ is the Levi-Civita symbol. All the repeated labels are summed. We see that the first term is the Laplacian for $g_{\mu\nu}$ applied to each component $f_a$. The last term is the cross term and the middle term is the Laplacian on $M_2(\C)$.  Here, at each $x,t$ (e.g. regarding $f$ as constant on spacetime),  
\begin{align*}
\Delta_{M_2}f&=(\ ,\ )\nabla((\del_i f) s^i))\\
&=(\del_k f)(\ ,\ )\nabla s^k+\partial_j\partial_i f (\ ,\ )s^j\tens s^i\\
&=(\del_k f)(\ ,\ )(E^k_{i\alpha 0}(\extd x^\alpha\tens s^i+ s^i\tens \extd x^\alpha)+\gamma^k_{ij} s^i\tens s^j)+(\partial_i\partial_j f) h^{ij}\\
&=f_c(\partial_i\partial_j \sigma^c+(\del_k \sigma^c)\gamma^k_{ij})h^{ij}.
\end{align*}
Using $\del_i\sigma^b=-\epsilon_{ibc}\sigma^c$ and $h^{11}=-h^{22}$,
\begin{align*}
\Delta_{M_2}f&=(f_i\sigma^j-\epsilon_{kcb} f_c\sigma^{b}\gamma^k_{ij})h^{ij}\\
&=(\gamma^1_{ij2}-\gamma^2_{ij1})h^{ij}f_3\sigma^0+(f_i h^{i1}+(\imath f_2 \gamma^1_{ij2}-f_3\gamma^2_{ij0}-\imath f_1\gamma^2_{ij2})h^{ij})\sigma^1\\
&\quad+(f_i h^{i2}+(f_3\gamma^1_{ij0}-\imath f_2\gamma^1_{ij1}+\imath f_1\gamma^2_{ij1})h^{ij})\sigma^2\\
&\quad+(f_1\gamma^2_{ij0}-f_2\gamma^1_{ij0}-\imath f_3\gamma^1_{ij1}-\imath f_3\gamma^2_{ij2})h^{ij}\sigma^3.
\end{align*}
\endproof

Explicitly, 
\begin{align*}
&\Delta_{M_2}\sigma^0=0,\\
&\Delta_{M_2}\sigma^1=(h^{11}-\imath\gamma^2_{ij2}h^{ij})\sigma^1+(h^{12}+\imath\gamma^2_{ij1}h^{ij})\sigma^2+\gamma^2_{ij0}h^{ij}\sigma^3,\\
&\Delta_{M_2}\sigma^2=(h^{21}+\imath\gamma^1_{ij2}h^{ij})\sigma^1+(h^{22}-\imath\gamma^1_{ij1}h^{ij})\sigma^2-\gamma^1_{ij0}h^{ij}\sigma^3,\\
&\Delta_{M_2}\sigma^3=(\gamma^1_{ij2}-\gamma^2_{ij1})h^{ij}\sigma^0-\gamma^2_{ij0}h^{ij}\sigma^1+\gamma^1_{ij0}h^{ij}\sigma^2-\imath(\gamma^1_{ij1}+\gamma^2_{ij2})h^{ij}\sigma^3.
\end{align*}
so that $\Delta_{M_2}$ as a matrix in the basis $\{\sigma^0,\sigma^1,\sigma^2,\sigma^3\}$ can be written as
\begin{equation*}
\Delta_{M_2} = 
\begin{pmatrix}
0 & 0 & 0 & 0 \\
0 & \ h^{11}-\imath\gamma^2_{ij2}h^{ij}\  & h^{12}+\imath\gamma^2_{ij1}h^{ij}\  & \ \gamma^2_{ij0}h^{ij} \\
0  & \ h^{21}+\imath\gamma^1_{ij2}h^{ij}\   & \ h^{22}-\imath\gamma^1_{ij1}h^{ij}\  & \ -\gamma^1_{ij0}h^{ij}  \\
(\gamma^1_{ij2}-\gamma^2_{ij1})h^{ij}\  & \ -\gamma^2_{ij0}h^{ij}\  & \ \gamma^1_{ij0}h^{ij}\  & \ -\imath(\gamma^1_{ij1}+\gamma^2_{ij2})h^{ij}\  
\end{pmatrix}.
\end{equation*}

These results are general. We now focus on the standard quantum metric on $M_2(\C)$ but allow the scale of this (and the parameter $\rho$ in its  QLC) to vary over spacetime. 

\begin{proposition}\label{hscaled} Let $h_{ij}s^i\tens s^j=h(x,t)(s^1\tens s^1-s^2\tens s^2)$ be a scaling of the standard  QRG of $M_2(\C)$ with its 1-parameter QLC in  Section~\ref{QRGM2}. Then 
\[E^2_{1\alpha 0}=E^1_{2\alpha 0}=\frac{\imath \rho(x,t)}{2h(x,t)} \partial_\alpha h(x,t),\quad E^1_{1\alpha 0}=E^2_{2\alpha 0}=-\frac{1}{2h(x,t)}\partial_\alpha h(x,t)\]
solve the conditions for a QLC on the product in Proposition~\ref{propA0}. Moreover, 
\begin{align*}
\Delta f(x,t)&=(\Delta_{LB} f_a(x,t))\sigma^a- {1\over h(x,t)}\left(-f_1(x,t)\sigma^1+f_2(x,t)\sigma^2-g^{\mu\nu}(\partial_\nu h(x,t)\partial_\mu f_a(x,t))\sigma^a\right). 
\end{align*}
If $h(x,t)$ is constant in spacetime and we let $h^{-1}(x,t)=\delta_m$ then the KG equation operator reduces to 
\begin{align*}
(\Delta +m^2)f&=(\Delta_{LB} f_a)\sigma^a+ m^2 f_0 + (m^2+\delta_m) f_1(x,t)\sigma^1+(m^2-\delta_m) f_2(x,t)\sigma^2+ m^2 f_3\sigma^3,
\end{align*}
which implies a splitting of the masses of the $f_1,f_2$ components of $f$ regarded a quadruplet of fields on $M$. Here $a\in\{0,1,2,3\}$.
\end{proposition}
\proof  In fact, we already described the 1-parameter QLC for this form of quantum metric on $M_2(\C)$ in Section~\ref{QRGM2} and the scale factor $h(x,t)$ in front does not change this. In terms of the general analysis above, the corresponding $\gamma$ tensor is 
\begin{align*}
&\gamma^1_{111}=\gamma^1_{221}=\gamma^2_{112}=\gamma^2_{222}=\gamma^1_{212}=\gamma^2_{121}=-\gamma^1_{122}=-\gamma^2_{211}=\frac{\imath}{2},\\
&\gamma^1_{121}=\gamma^2_{212}=\gamma^1_{112}=\gamma^2_{111}=\gamma^1_{222}=\gamma^2_{221}=\gamma^i_{jk0}=0,\\
&\gamma^1_{211}=\gamma^2_{122}=-\rho(x,t),\quad \gamma^k_{ij0}=0. 
\end{align*}
and results in 
\begin{equation*}
\Delta_{M_2} = 
\begin{pmatrix}
0 & 0 & 0 & 0 \\
0 & \ h^{-1}(x,t)\  & 0 & 0 \\
0  & 0  & \ -h^{-1}(x,t)\  & 0\\
0  & 0 & 0 &0
\end{pmatrix}
\end{equation*} in line with Section~\ref{QRGM2}. Here $\rho$ is also allowed in principle to depend on spacetime but does not enter directly into the Laplacian. Then $\Delta f(x,t)$ in Theorem~\ref{thmA0} reduces as stated.
\endproof

Note that if we use the full 3-parameter connection from Section~\ref{QRGM2} then only $\Delta_{M_2}$ changes, now to (\ref{delta2m}). This gives effective square masses $m^2+\delta_m^i$ with
\begin{align*}
 &\delta_m^1=\frac{(u+v-2)}{4\rho h}(v-u-2\rho),\\
 &\delta_m^2=\frac{(u+v-2)}{4\rho h}(v-u+2\rho),\\
 & \delta_m^3=\frac{(u+v-2)}{2\rho h}(v-u)
\end{align*}
for the field $f_i$ in terms of the parameters $u,v,\rho$, instead of the formulae above. These are again in principle allowed to vary in spacetime, but the insteresting case is for constant shifts in square mass.  Finally, we compute the curvatures of the tensor product quantum geometry in the 1-parameter case (the formulae for the 3-parameter case are more complicated and are omitted).

\begin{theorem}\label{thmSA0} The Ricci curvature of the QLC on the product in Proposition~\ref{propA0}  is 
\[ S=S_M+S_{M_2}+\frac{1}{2}g^{\alpha\beta}\left(E^n_{m\alpha 0}E^n_{m\beta 0}-h_{nm}h^{nm}(\nabla_\alpha E^i_{i\beta 0}-E^i_{j\alpha 0}E^j_{i\beta 0})\right),\]
where $S_M$ is the classical Ricci scalar for $g_{\mu\nu}$ in our conventions and
\[ S_{M_2}=-\frac{1}{2}h^{jk}(\gamma^k_{mn}\gamma^n_{mj}-\imath \gamma^k_{mj}\sigma^m+\gamma^k_{mjb}\epsilon_{mbc}\sigma^c)\]
is the QRG scalar curvature on $M_2(\C)$ at each $x,t$. Here $n,m,i,j,k\in\{1,2\},b\in\{1,2\},c\in\{1,2,3\}$.
\end{theorem}
\proof  We first compute the Riemann curvature for the connection~Proposition~\ref{propA0},
\begin{align*}
R_\nabla \extd x^\mu&=-(\partial_\nu\Gamma^\mu_{\alpha\beta}+\Gamma^\mu_{\nu k}\Gamma^k_{\alpha\beta})\extd x^\nu\wedge\extd x^\alpha\tens \extd x^\beta+(\partial_\nu C^\mu_{ij}+\Gamma^\mu_{\nu\beta}C^\beta_{ij}+C^\mu_{im}E^m_{j\nu 0})\extd x^\nu \wedge s^i\tens s^j\\
&-(C^\mu_{ji}\sigma^j+\imath C^\mu_{mj}\gamma^j_{mi}){\rm Vol}\tens s^i-\imath C^\mu_{mj}E^j_{m\alpha 0}{\rm Vol}\tens\extd x^\alpha,\\
R_\nabla s^k&=(\partial_\mu E^k_{i\nu 0}-E^k_{j\mu 0}E^j_{i\nu 0})\extd x^\mu\wedge\extd x^\nu\tens s^i+(\partial_\mu E^k_{i\nu 0}-E^k_{j\mu 0}E^j_{i\nu 0}-E^k_{i\alpha 0}\Gamma^\alpha_{\mu\nu})\extd x^\mu\wedge s^i\tens\extd x^\nu\\
&-(E^k_{i\mu 0}\sigma^i+\imath E^j_{m\mu 0}\gamma^k_{mj}){\rm Vol}\tens \extd x^\mu+((\partial_\mu \gamma^k_{ija})\sigma^a+\gamma^k_{im}E^m_{j\mu 0})\extd x^\mu\wedge s^i\tens s^j\\
&-(\gamma^k_{ji}\sigma^j+\imath\gamma^k_{mj}\gamma^j_{mi}+\imath\gamma^k_{mib}\epsilon_{mbc}\sigma^c+\imath E^k_{m\alpha 0}C^\alpha_{mi}){\rm Vol}\tens s^i.
\end{align*}
This then leads to the Ricci tensor,
\begin{align*}
&{\rm Ricci}=\frac{1}{2}\left(-\partial_\mu\Gamma^\mu_{\alpha\beta}-\Gamma^\mu_{\mu k}\Gamma^k_{\alpha\beta}+\partial_\alpha\Gamma^\mu_{\mu\beta} +\Gamma^\mu_{\alpha k}\Gamma^k_{\mu\beta}-h_{nk}h^{ni}(\partial_\alpha E^k_{i\beta 0}-E^k_{j\alpha 0}E^j_{i\beta 0}-E^k_{i\mu 0}\Gamma^\mu_{\alpha\beta})\right)\extd x^\alpha\tens \extd x^\beta\\
&\quad+\frac{1}{2}\imath h_{nk}h^{ni}(E^k_{j\mu 0}\sigma^j+\imath E^j_{m\mu 0}\gamma^k_{mj})s^i\tens \extd x^\mu-\frac{1}{2}h_{nk}h^{ni}((\partial_\mu\gamma^k_{ija})\sigma^a+\gamma^k_{im}E^m_{j\mu 0})\extd x^\mu\tens s^j\\
&\quad+\frac{1}{2}(g^{\mu\alpha}h_{mj}(\Gamma^\beta_{\mu\alpha}E^m_{i\beta 0}+E^m_{n\mu 0}E^n_{i\alpha 0}-\partial_\mu E^m_{i\alpha 0})\\
&\quad+\imath h_{pk}h^{pi}(\gamma^k_{mj}\sigma^m+\imath\gamma^k_{mn}\gamma^n_{mj}+\imath\gamma^k_{mjb}\epsilon_{mbc}\sigma^c-\imath g^{\alpha\beta}h_{nj}E^k_{m\alpha 0}E^n_{m\beta 0}))s^i\tens s^j.
\end{align*}
Finally, we apply $(\ ,\ )$ to obtain the Ricci scalar $S$,
\begin{align*}
S&=(\ ,\ ){\rm Ricci}=\frac{1}{2}g^{\alpha\beta}(-\partial_\mu\Gamma^\mu_{\alpha\beta}-\Gamma^\mu_{\mu k}\Gamma^k_{\alpha\beta}+\partial_\alpha\Gamma^\mu_{\mu\beta} +\Gamma^\mu_{\alpha k}\Gamma^k_{\mu\beta})\\
&+\frac{1}{2}g^{\alpha\beta}(h_{pk}h^{pi}h^{ij}h_{nj}E^k_{m\alpha 0}E^n_{m\beta 0}-(h_{nk}h^{ni}+h_{kn}h^{in})(\partial_\alpha E^k_{i\beta 0}-E^k_{j\alpha 0}E^j_{i\beta 0}-E^k_{i\mu 0}\Gamma^\mu_{\alpha\beta}))\\
&+\frac{1}{2}\imath h_{pk}h^{pi}h^{ij}(\gamma^k_{mj}\sigma^m+\imath\gamma^k_{mn}\gamma^n_{mj}+\imath\gamma^k_{mjb}\epsilon_{mbc}\sigma^c)\\
&=\frac{1}{2}g^{\alpha\beta}(-\partial_\mu\Gamma^\mu_{\alpha\beta}+\Gamma^\mu_{\mu k}\Gamma^k_{\alpha\beta}+\partial_\alpha\Gamma^\mu_{\mu\beta} -\Gamma^\mu_{\alpha k}\Gamma^k_{\mu\beta})\\
&+\frac{1}{2}g^{\alpha\beta}( E^n_{m\alpha 0}E^n_{m\beta 0}-h_{nm}h^{nm}(\partial_\alpha E^i_{i\beta 0}-E^i_{j\alpha 0}E^j_{i\beta 0}-E^i_{i\mu 0}\Gamma^\mu_{\alpha\beta}))\\
&+\frac{1}{2}\imath h^{jk}(\gamma^k_{mj}\sigma^m+\imath\gamma^k_{mn}\gamma^n_{mj}+\imath\gamma^k_{mjb}\epsilon_{mbc}\sigma^c).
\end{align*}
Here we used $h_{pk}h^{pi}h^{ij}h_{nj}=-h_{pk}\frac{h_{pi}}{\det(h)}h^{ij}h_{nj}=-h_{pk}\frac{\delta^j_{p}}{\det(h)}h_{nj}=-\frac{h_{jk}h_{nj}}{\det(h)}=h_{jk}h^{nj}=\delta^n_k$ and $h_{nk}h^{ni}+h_{kn}h^{in}=h_{nm}h^{nm}\delta^i_k$. We recognise the first and last terms as respectively the spacetime curvature (which in our conventions $-{1\over 2}$ of the usual value) and an expression which can be identified as the scalar curvature $S_{M_2}$ for the QRG on $M_2(\C)$ defined by $\gamma$. \endproof

We use $\nabla_\mu,\nabla^\mu=g^{\mu\nu}\nabla_\nu$ for the usual Levi-Civita covariant derivative for the metric $g_{\mu\nu}$. These results are again general and we now focus on the scaled standard QRG on $M_2(\C)$.

\begin{proposition}\label{propShscaled} Let $h_{ij}s^i\tens s^j=h(x,t)(s^1\tens s^1-s^2\tens s^2)$ as in Proposition~\ref{hscaled}. If we assume that $\rho(x,t)$ is constant and let $ h(x,t)=e^{\frac{2}{\sqrt{3-3\rho^2}}\phi(x,t)}$, then
\[ S=S_M+ \nabla^\mu \phi\nabla_\mu \phi+\frac{2}{\sqrt{3-3\rho^2}}\nabla^\mu\nabla_\mu \phi\]
where  $\phi(x,t)$ is real scalar field and $\rho$ is now an imaginary parameter. Moreover,
\[E^2_{1\mu 0}=E^1_{2\mu 0}=\frac{\imath \rho}{\sqrt{3-3\rho^2}} \partial_\mu \phi(x,t),\quad E^1_{1\mu 0}=E^2_{2\mu 0}=-\frac{1}{\sqrt{3-3\rho^2}}\partial_\mu \phi(x,t).\] 
\end{proposition}
\proof  Substituting $\gamma $s in Section~\ref{QRGM2} and $E$ in Proposition~\ref{hscaled} into $S$, we find 
\begin{align*}
S&=S_M+\frac{1}{2}g^{\alpha\beta}\left(\frac{3-3\rho^2}{2}(\partial_\alpha lnh)(\partial_\beta lnh)+2\partial_\alpha\partial_\beta lnh-2\Gamma^\mu_{\alpha\beta}(\partial_\mu lnh)\right)+0\\
&=S_M+\frac{1}{2}g^{\alpha\beta}\left(\frac{3-3\rho^2}{2}(\partial_\alpha lnh)(\partial_\beta lnh)+2\nabla_\alpha\partial_\beta lnh\right)\\
&=S_M+\frac{3-3\rho^2}{4}(\nabla^\alpha lnh)(\nabla_\alpha lnh)+\nabla^\alpha\nabla_\alpha lnh,
\end{align*}
where $\nabla_\alpha,\nabla^\alpha=g^{\alpha\beta}\nabla_\beta$. 

Next, under the reality condition on the quantum metric, we know $h(x,t)$ is real, and under $*$-preserving condition on the connection, $\rho(x,t)$ is an imaginary functional parameter so that $3-3\rho^2\ge 3$ is real. Hence, if $\rho$ is constant then we can let $h(x,t)=e^{\frac{2}{\sqrt{3-3\rho^2}}\phi(x,t)}$ and then we find $S$ and $E$ as stated. \endproof

The last term of $S$ is a total divergence and hence will vanish on integration over $M$ when we look at the corresponding Einstein-Hilbert action. This therefore shows the same mechanism for the dynamical perturbation of mass as in \cite{ArgMa4} in which a similar field to $\phi$ appears as a real scalar field and enters as in Proposition~\ref{hscaled} to produce a  square mass variation
\begin{equation}\label{deltamphi} \delta_m=e^{-\frac{2}{\sqrt{3-3\rho^2}}\phi(x,t)}.\end{equation}

\section{Alternate QRG on $M_2(\C)$}\label{secalt}

Here we consider another choice of quantum metric for which the QRG is well understood, namely $\cg=\imath (s^2\tens s^1-s^1\tens s^2)$  on $M_2(\C)$ as in \cite{BegMa,LirMa3}. The physical interpretation here is less clear so we will give this only briefly.  A natural 4-parameter QLC here was found in \cite{BegMa} and in our $s^i$ basis becomes
 \begin{align*}
 &\gamma^1_{221}=\imath-\gamma^1_{111},\quad \gamma^2_{121}=2\imath-\gamma^1_{111},\quad \gamma^2_{211}=\imath-\frac{\imath}{2}(a+b+u+v)-\gamma^1_{111},\\
 &\gamma^1_{212}=\gamma^1_{111}-\frac{\imath (a-b)(av-bu)}{4ab},\quad \gamma^2_{222}=2\imath-\gamma^1_{212},\quad \gamma^2_{112}=\gamma^1_{212}-\imath,\\
  &\gamma^1_{122}=\gamma^1_{212}-\imath+\frac{\imath}{2}(a+b-u-v),\quad \gamma^1_{211}=-\gamma^2_{221}=\gamma^2_{111},\quad \gamma^1_{121}=\gamma^2_{111}+\frac{1}{2}(a-b-u+v),\\
  &\gamma^1_{112}=\gamma^2_{111}-\frac{(u+v)(bu-av)}{4ab},\quad \gamma^1_{222}=\gamma^2_{122}=-\gamma^1_{112},\quad \gamma^2_{212}=-\gamma^1_{112}+\frac{1}{2}(a-b+u-v),\\
  &\gamma^1_{111}=\imath\left(1+\frac{(av+bu-2ab)(a+b+u+v)}{8ab}\right),\quad \gamma^2_{111}=\frac{(bu-av)(a+b+u+v)}{8ab}.
 \end{align*}
The reality condition for a $*$-preserving connection  is $v=-\bar u,b=-\bar a$. We now calculate the Ricci curvature, Ricci tensor and Ricci scalar. A further restriction that avoids matrices in the Ricci scalar is  $v=-u$ and hence $u$ real. In this case,
\begin{align*}
R_{\nabla}s^i&=(-1)^i\frac{\imath}{2}(a+b){\rm Vol}\tens s^i+\frac{1}{2}(a-b+(-1)^i 2u){\rm Vol}\tens s^{\bar i},\\
{\rm Ricci}&=\frac{1}{4}((a+b)s^1\tens s^1+\imath (a-b-2u)s^1\tens s^2+\imath (a-b+2u)s^2\tens s^1-(a+b)s^2\tens s^2),\\
S&=-u.
\end{align*}
 The Laplacian $\Delta_{M_2}$ is 
\begin{align*}
\Delta_{M_2}f&=-(2+u+v)f_3\sigma^0+{1\over 2}((a-b+u-v) f_1 +\imath (a+b-u-v)f_2)\sigma^1\\
&+{1\over 2}(\imath (a+b+u+v)f_1-(a-b-u+v)f_2)\sigma^2+(u-v)f_3\sigma^3\\
=&-2f_3\sigma^0+{1\over 2}((a-b+2u) f_1 +\imath (a+b)f_2)\sigma^1+{1\over 2}(\imath (a+b)f_1-(a-b-2u)f_2)\sigma^2+2u f_3\sigma^3
\end{align*}
in the case $v=-u$ and $u$ real. The 1-parameter case of interest further sets $a=b=0$ so that 
\[ \nabla s^i=\imath(\sigma^1 s^1+\sigma^2 s^2)\tens s^i-u\epsilon_{ik}\sigma^i s^i\tens s^k, \quad \Delta_{M_2}f=uf_1\sigma^1+uf_2\sigma^2+2uf_3\sigma^3-2f_3\sigma^0\]
as one can then check directly for this quantum metric. Note that $\Delta_{M_2}$ has $f_0$ as a zero mode, and also the action ${1\over 2}{\rm Tr}(\bar f\Delta_{M_2} f)$ is not real due to a $\bar f_0 f_3$ term. 

If we put this quantum metric into the general setting of Lemma~\ref{lem_metric} then (after a long calculation) one can show that we are forced to $A=0$, as in the preceding section. In this case we find from Theorem~\ref{thmSA0} that

\begin{proposition} If  $h_{ij}s^i\tens s^j=\imath h(x,t)(s^2\tens s^1-s^1\tens s^2)$ and we set  $h(x,t)=e^{2\phi(x,t)}$ and $\gamma$ to be the 1-parameter QLC with $v=-u,a=b=0$ then
\[ S=S_M-\nabla_\mu\phi\nabla^\mu\phi+2\nabla_\mu\nabla^\mu \phi-u\]
where  $\phi(x,t)$ is real scalar field. Here $u$ is allowed to vary in spacetime.
\end{proposition}
\proof 
Conditions (\ref{he}) and (\ref{heg}) in Proposition~\ref{propA0} for $h(x,t)=e^{2\phi(x,t)}$ become \[E^2_{1\alpha 0}=E^1_{2\alpha 0}=\imath E^1_{1\alpha 0}+\imath\partial_\alpha \phi(x,t),\quad E^2_{2\alpha 0}=-E^1_{1\alpha 0}-2\partial_\alpha \phi(x,t)\]
with $E^1_{1\alpha 0}$ arbitrary. 
Substituting $E$ and $\gamma$ into $S$, we obtain
\begin{align*}
S&=S_M-u+\frac{1}{2}g^{\alpha\beta}(-2\partial_\alpha\phi\partial_\beta\phi+4\nabla_\alpha\partial_\beta \phi)\end{align*}
which we write as stated. 
\endproof
The last term of $S$ is the Ricci scalar of $M_2(\C)$ and the term before that is a total divergence and can be dropped when we integrate over $M$. Hence we have the same kind of result as before, namely the scaling field $h$ appears as a scalar field $\phi$, but note that its action enters with a negative sign. On the other hand, Theorem~\ref{thmA0} now gives 
\[ \Delta f =\Delta_{LB} f  +\frac{1}{h(x,t)}(uf_1\sigma^1+uf_2\sigma^2+2uf_3\sigma^3-2f_3\sigma^0-\partial^\mu h(x,t) \partial_\mu f_a\sigma^a ).\]
In the simplest case where $h$ is a constant, the scalar field action $S_f= \int \bar f\Delta f$ using ${1\over 2}{\rm Tr}$ for the integration over $M_2$ is then
\[S_f=\int_M\sqrt{-g}\prod \extd x^\mu\left(\sum_a \bar f_a\Delta_{LB}f_a +e^{-2\phi}(-2 \bar f_0 f_3+ u \bar f_1 f_1 + u\bar f_2 f_2+ 2u \bar f_3 f_3 )\right),\]
showing a change of mass of the $f_a$ modes for $a=1,2,3$ and an unphysical coupling of the $f_0,f_3$ modes. The latter can be avoided if we restrict to a triplet of fields with $f_0=0$. 

\section{Concluding remarks}

We have shown how the assumption of a finite noncommutative geometry, in our case with `coordinate algebra' $M_2(\C)$ adjoined to classical spacetime $M$ (by a tensor product of the coordinate algebras) leads to two different physical phenomena. One solution for the QRG on the product, in  Section~\ref{secA0}, behaves remarkably similarly to \cite{ArgMa4} where we adjoined a completely different finite QRG, namely a discrete circle $\Z_n$. In both cases, the finite QRG metric is allowed to vary over spacetime by a scale factor $h(x,t)$ and its logarithm enters the scalar curvature (and hence the Einstein-Hilbert action) for the product QRG as a massless real scalar field field $\phi$, i.e. becomes a dynamic field, see Proposition~\ref{propShscaled}. Moreover, the mass coupling to scalar field multiplets when the KG equation on the product QRG is broken down in terms of component fields on $M$, enters as the exponential of $\phi$ in the Lagrangian, see (\ref{deltamphi}), so that even a ground state $\phi=0$ produces mass. Also, in both cases the quantum metric on the product does  not have cross terms due to natural restrictions coming from the QRG. This confirms, for a second independent model, a novel mechanism of dynamical mass-splitting (or dynamical mass generation) induced by tensoring with a finite noncommutative geometry.

Our other phenomenon, as in Proposition~\ref{propsol+}, is a complementary QRGs on the product and is closer to the spirit of the original Kaluza-Klein hypothesis in that cross terms in the quantum metric {\em are} allowed and the scale function $h$ for the $M_2(\C)$ geometry is instead forced to be a constant. The cross terms correspond to a 1-form $A_\mu$ on the spacetime $M$ and a real scalar field appears as a multiplet where two of the real components combine to a single charged scalar field $\psi$ minimally coupled to $A_\mu$. The latter is again a dynamic field in that it enters the scalar curvature on the product (and hence the Einstein-Hilbert action), but this time with a Maxwell action $||F||^2$, see Theorem~\ref{thmS} but with an additional term $||A.F||^2$. Hence this is a dynamical charge generation phenomenon where again part of the geometry on the product appears as a new field on $M$. This is not unlike the basic discovery of Kaluza-Klein but surprising to see it here for a finite QRG as the extra factor rather than $S^1$.

There are a great many questions that arise naturally from the present work. One of these, but not at all easy to address, would be to analyse the automorphisms of the QRG on the product and see to what extent these reduce to diffeomorphisms on $M$ and $U(1)$ gauge transformations such that $A_\mu$ transforms as expected. This requires a general theory of such automorphisms which would first need to be developed. Here any algebra has a `measuring bialgebra' of all possible transformations, which should be cut down to respect the differential calculus, etc. In the paper, the $A_\mu$ arises very differently as a 1-form entering into the quantum metric and given terms such as $||A.F||$, if a gauge potential arises from the theory (as the minimal coupling to a scalar field suggests) then it does so as intrinsically gauge fixed. Which gauge is not at all clear but a natural simplification suggested by Lemma~\ref{3eq} was to impose the Killing form equation $\tilde\nabla_{(\mu}A_{\nu)}=0$  as a stronger version of  Lorentz gauge $\tilde g^{\alpha\beta}\tilde\nabla_\alpha A_\beta=0$. With this assumption, $A.F=-\extd ||A||^2$ and the equations of motion for $A$ including the $||A.F||^2$ term are 
\[ \tilde g^{\alpha\beta}\tilde\nabla_\alpha F_{\beta\mu}=J_\mu,\quad   J_\mu=\tilde g^{\alpha\beta}\del_\alpha(||A||^2)F_{\beta\mu}.\]
The effective `source' here is not automatically conserved.  Even this special case, as well as the full equations of motion, remain to be better understood. We can, however, make one remark, which is the sense in which the $||A.F||$ term is a perturbation. Naively, this would seem to be at large $h$, but in fact this is misleading when we match the $||F||^2$ to Maxwell theory.

First off, putting in physical constants but with $c=1$, we suppose that the Einstein-Hilbert action on the product is
\[ -{1\over 8\pi G}\int_M \sqrt{-\tilde g} \prod\extd x^\mu\,  S\]
remembering that our $S$ is $-1/2$ of the usual conventions for the scalar curvature. This then recovers the correct Einstein-Hilbert action for $\tilde g$  with the signature (- + + +) from the first term of $S$ in Theorem~\ref{thmS}. Note that $S$ has units of length${}^{-2}$ and hence $h$ has units of square-length. Moreover, $A_\mu/h$ in Proposition~\ref{proplap} needs to have length${}^{-1}$ so $A_{\mu}$ as units of length. Moreover, comparing with the covariant derivative for a particle of charge $e$, Maxwell potential $\bar A_\mu$ and curvature $\bar F$, we need 
\[ A_\mu = {e h\over\hbar}\bar A_\mu,\quad ||F||^2= {e^2 h^2\over \hbar^2}||\bar F||^2={4\pi\alpha\eps_0}{h^2\over \hbar}||\bar F||^2\]
where $\alpha$ is the fine structure constant if we take $e$ to be the electron charge, say. If we want the above action to recover
\[- {\eps_0 \over 4}\int_M \sqrt{-\tilde g} \prod\extd x^\mu\, ||\bar F||^2\]
from the third term of $S$ in Theorem~\ref{thmS}, then we need
\[ h = { 4 G\hbar  \over \alpha  }= {4\lambda_P^2\over\alpha}\]
where $\lambda_P$ is the Planck length. In this case, the last two terms in $S$ with physical fields and constants are proportional to 
\[ ||\bar F||^2 + {e^2 h\over \hbar^2}||\bar A.\bar F||^2= ||\bar F||^2 +   16\pi G \eps_0 ||\bar A.\bar F||^2   \]
hence this second term is generically suppressed if  the 1-form field strength is weak in the rough estimate sense
\[ ||\bar A||<< {1\over 4 \sqrt{\pi G \eps_0}}.\]
Of course, this is not something one can properly consider in gauge theory as this quantity is not gauge-invariant but, as mentioned above, the theory should presumably be compared with Maxwell theory in a particular gauge. Ignoring this for purposes of discussion, to reach this bound for the field strength $||\bar A||\sim {|e|\over 4\pi \eps_0 r}$ at radius $r$ from a classical electron point source, we would need to be within  2$\sqrt{\alpha}\approx {1/6}$ Planck lengths of the origin, i.e. in practice this would not be reached due to quantum gravity effects. One can conversely speculate that the $||A.F||^2$ term could prevent theoretical issues such as the classically infinite self-energy of the electron, but this remains to be studied further.

Another direction for further work would be to follow the programme of this paper and of \cite{ArgMa4} but for other  QRGs tensored onto spacetime. Indeed,  the second case with $A_\mu$ did not result in any modified mass to a scalar field because we were forced to the case where the Laplacian $\Delta_{M_2}=0$ on $M_2(\C)$ (see Proposition~\ref{propsol+}), which is rather special. Likewise in Section ~\ref{secA0}, natural restrictions on the QRG led to $A_\mu=0$ but these could be relaxed. One could also look at other metrics on $M_2(\C)$ as well as other algebras entirely.  Here, much of the analysis for the fuzzy sphere here in this role as in \cite{ArgMa4} but the equations for a QLC on the product were too hard to solve. This could certainly be pursued further. Similarly, one could put a chosen finite graph QRG as each point of spacetime, where the QRG of the $A_n$ graph $\bullet$-$\bullet$-$\cdots$-$\bullet$ was recently solved in \cite{ArgMa3}. It would also be interesting to consider quantum geodesics on products and relate them to physics on $M$. These are a relatively new concept but already give a new point of view in the classical case on the Ricci curvature \cite{BegMa:cur}. The quantum geodesic formalism was recently elaborated for a particular quantum Minkowski spacetime model in \cite{LiuMa}. 

 Finally, it would be critical for physics to extend our analysis for the KG equation on the product in this paper to the Dirac equation. Apart from obviously being needed to include fermionic matter fields, this could also make contact with Connes approach to the standard model, e.g. \cite{Con1,Cham}, where a finite noncommutative geometry is again tensored onto classical spacetime. A Connes spectral triple already models a Dirac operator, so this aspect is already covered. Also, the automorphisms can be analysed and, in the spectral action, a Higgs field can be seen to emerge. Coming at the same problem within QRG is much harder as we do not necessarily even have a Dirac operator and therefore have to build one from a spinor bundle, bimodule connection and other data compatible with the quantum metric. However, one can indeed arrive at Connes spectral triples or something similar in several known cases, e.g. \cite{BegMa:spe, LirMa3,Ma:dir} allowing for an intersection of the two approaches. The potential here is that this additional restriction (that a spectral triple be realisable within QRG) could end up explaining features of particle physics which might not be explained if we allowed any spectral triple.  These are some key directions for further work.

 \end{document}